\pdfoutput=1

\newif\ifAOAS
\AOAStrue      

\ifAOAS
    \makeatletter
    \def\input@path{{preamble_aoas/}}
    \makeatother
\fi

\ifAOAS
    \documentclass[aoas,preprint]{imsart}
\else
    \documentclass[12pt]{article}
\fi

\usepackage{amsmath}
\usepackage{amssymb}
\usepackage{mathtools}
\usepackage{graphicx}
\usepackage{booktabs}
\usepackage{array}
\usepackage{rotating}
\usepackage{xspace}
\usepackage{todonotes}
\usepackage{dsfont}
\usepackage{multirow}
\usepackage{float}

\usepackage[authoryear]{natbib}

\usepackage[colorlinks,linktoc=all]{hyperref}
\usepackage[nameinlink,capitalise]{cleveref}

\newcommand{\E}[2]{\mathbb{E}_{#1}\left[#2\right]}

\newcommand{\defeq}{\overset{\mathrm{def}}{=}}

\newcommand{\R}[0]{\mathbb{R}}

\newcommand{\normal}{\mathcal{N}}

\newcommand{\paren}[1]{\left(#1\right)}

\newcommand{\modelname}{GrAFT-PDX\xspace}

\ifAOAS
    \input{preamble_aoas}
\else

\usepackage[utf8]{inputenc} 
\usepackage[T1]{fontenc}    

\usepackage[bitstream-charter]{mathdesign}
\usepackage{amsmath}
\usepackage[scaled=0.92]{PTSans}

\usepackage[
  paper  = letterpaper,
  left   = 1.0in,
  right  = 1.0in,
  top    = 1.0in,
  bottom = 1.0in,
  ]{geometry}

\usepackage[usenames,dvipsnames,table]{xcolor}
\definecolor{shadecolor}{gray}{0.9}

\usepackage[final,expansion=alltext]{microtype}
\usepackage[english]{babel}
\usepackage[parfill]{parskip}
\usepackage{afterpage}
\usepackage{framed}

{\endMakeFramed}

\usepackage{lineno}

\usepackage{ragged2e}

\newcounter{parcount}

\usepackage{graphicx}
\usepackage{wrapfig}
\usepackage[labelfont=bf,font=footnotesize,width=.9\textwidth]{caption}
\usepackage[format=hang]{subcaption}

\usepackage{booktabs,multirow,multicol}       

\usepackage[algoruled]{algorithm2e}
\usepackage{listings}
\usepackage{fancyvrb}
\fvset{fontsize=\normalsize}

\usepackage[affil-it]{authblk}

\usepackage[colorlinks,linktoc=all]{hyperref}
\usepackage[all]{hypcap}
\hypersetup{citecolor=MidnightBlue}
\hypersetup{linkcolor=MidnightBlue}
\hypersetup{urlcolor=MidnightBlue}

\usepackage[nameinlink,capitalise]{cleveref}
\Crefname{section}{\S}{\S}

\usepackage[acronym,nowarn]{glossaries}

\lstdefinestyle{mystyle}{
    commentstyle=\color{OliveGreen},
    keywordstyle=\color{BurntOrange},
    numberstyle=\tiny\color{black!60},
    stringstyle=\color{MidnightBlue},
    basicstyle=\ttfamily,
    breakatwhitespace=false,
    breaklines=true,
    captionpos=b,
    keepspaces=true,
    numbers=left,
    numbersep=5pt,
    showspaces=false,
    showstringspaces=false,
    showtabs=false,
    tabsize=2
}
\usepackage{centernot}
\usepackage{amsthm}         
\usepackage{nicefrac}       
\usepackage{mathtools}      
\usepackage{amsbsy}         
\usepackage{amstext}        
\usepackage{thmtools}       
\usepackage{thm-restate}    

\begingroup
    \makeatletter
    \@for\theoremstyle:=definition,remark,plain\do{%
        \expandafter\g@addto@macro\csname th@\theoremstyle\endcsname{%
            \addtolength\thm@preskip\parskip
            }%
        }
\endgroup

\crefname{lemma}{lemma}{lemmas}
\Crefname{lemma}{Lemma}{Lemmas}
\crefname{thm}{theorem}{theorems}
\Crefname{thm}{Theorem}{Theorems}
\crefname{prop}{proposition}{propositions}
\Crefname{prop}{Proposition}{Propositions}
\crefname{assumption}{assumption}{assumptions}
\crefname{assumption}{Assumption}{Assumptions}

\usepackage{arydshln}
\makeatletter
\def\adl@drawiv#1#2#3{%
        \hskip.5\tabcolsep
        \xleaders#3{#2.5\@tempdimb #1{1}#2.5\@tempdimb}%
                #2\z@ plus1fil minus1fil\relax
        \hskip.5\tabcolsep}
\newcommand{\cdashlinelr}[1]{%
  \noalign{\vskip\aboverulesep
           \global\let\@dashdrawstore\adl@draw
           \global\let\adl@draw\adl@drawiv}
  \cdashline{#1}
  \noalign{\global\let\adl@draw\@dashdrawstore
           \vskip\belowrulesep}}
\makeatother

\renewcommand{\epsilon}{\varepsilon}

\declaretheorem[style=plain,numberwithin=section,name=Theorem]{theorem}

\declaretheorem[style=definition,sibling=theorem,name=Example]{example}

\newenvironment{example*}
 {\pushQED{\qed}\example}
 {\popQED\endexample}
\numberwithin{equation}{section}

\definecolor{WowColor}{rgb}{.75,0,.75}
\definecolor{SubtleColor}{rgb}{0,0,.50}

\newcounter{margincounter}

\fi

\begin{document}

\ifAOAS
    \begin{frontmatter}

\title{\modelname: Growth and Failure-Time Modeling\\of Patient-Derived Xenograft Experiments}

\runtitle{\modelname}

\begin{aug}
\author[A]{\fnms{Nihar}~\snm{Mauskar}\ead[label=e1]{niharm@uchicago.edu}}
\author[B]{\fnms{Sebastian}~\snm{Salazar}}
\author[C,D]{\fnms{Aaron}~\snm{Schein}\ead[label=e2]{schein@uchicago.edu}}
\author[E]{\fnms{Wesley}~\snm{Tansey}\ead[label=e3]{tanseyw@mskcc.org}}

\address[A]{Department of Computer Science, University of Chicago\printead[presep={,\ }]{e1}}
\address[B]{Department of Computer Science, Columbia University}
\address[C]{Data Science Institute, University of Chicago\printead[presep={,\ }]{e2}}
\address[D]{Department of Statistics, University of Chicago}
\address[E]{Computational Oncology, Memorial Sloan Kettering Cancer Center\printead[presep={,\ }]{e3}}
\end{aug}

\begin{abstract}
Cancer drugs are often evaluated for efficacy in mice before proceeding to human clinical trials. To do this, human tumors are engrafted onto immune-incompetent mice to create a patient-derived xenograft (PDX). Engrafted tumors can be propagated to evaluate multiple drugs, but the time and costs involved in PDX studies limit high throughput drug screens in mice. Here we present \modelname, a Bayesian model of PDX drug screens that aims to reduce the PDX screening burden by enabling imputation of missing experiments and early stopping of likely unsuccessful drugs. Unlike conventional survival or growth models, \modelname jointly predicts tumor growth and survival outcomes using a learned latent function decomposition. In experiments on a large-scale PDX dataset, \modelname outperforms Cox proportional hazards models at predicting time of death, outperforms machine learning baselines at predicting tumor growth, and recovers known gene-treatment relationships without genomic input. 

\end{abstract}


\end{frontmatter}
\else
    \title{BayesPDX: Joint Tumor and Survival Prediction for Patient Derived Xenograft Data}

\author[1]{Nihar Mauskar}
\author[2]{Sebastian Salazar}
\author[3,4]{Aaron Schein}
\author[5]{Wesley Tansey}

\affil[1]{Department of Computer Science, University of Chicago}
\affil[2]{Department of Computer Science, Columbia University}
\affil[3]{Data Science Institute, University of Chicago}
\affil[4]{Department of Statistics, University of Chicago}
\affil[5]{Department of Computational Oncology, Memorial Sloan Kettering Cancer Center}

\date{}

\maketitle

\begin{abstract}
    Cancer drugs are often evaluated for efficacy in mice before proceeding to human clinical trials. To do this, human tumors are engrafted onto immune-incompetent mice to create a patient-derived xenograft (PDX). Engrafted tumors can be propagated to evaluate multiple drugs, but the time and costs involved in PDX studies limit high throughput drug screens in mice. Here we present \modelname, a Bayesian model of PDX drug screens that aims to reduce the PDX screening burden by enabling imputation of missing experiments and early stopping of likely unsuccessful drugs. Unlike conventional survival or growth models, \modelname jointly predicts tumor growth and survival outcomes using a learned latent function decomposition. In experiments on a large-scale PDX dataset, \modelname outperforms Cox proportional hazards models at predicting time of death, outperforms machine learning baselines at predicting tumor growth, and recovers known gene-treatment relationships without genomic input. 

\end{abstract}
\fi

\section{Introduction}
\label{sec:introduction}
\label{sec:intro}

Patient-derived xenografts (PDXs) provide an important method for studying how human tumors may respond to drug treatments in a controlled setting~\citep{tentler:etal:2012:pdx-cancer-review}. In these experiments, tumor tissue obtained from a cancer patient is implanted into an immunocompromised mouse. This allows researchers to monitor tumor growth over time \emph{in vivo}.  Different drugs or drug combinations can then be administered to individual mice to determine which therapies would more likely be effective in patients with similar tumors. Effective treatments shrink the size of the tumor, or slow its growth, and extend the overall survival of the mouse. When conducted across many mice engrafted with multiple PDXs, these experiments yield datasets consisting of longitudinal tumor measurements coupled with mouse survival outcomes.

Because each experiment measures the response of a specific tumor to a particular drug, these datasets let statistical models learn how tumors respond to treatment and identify promising therapies directly from the data. A PDX line can be propagated indefinitely, so the same tumor can be screened against many drugs. PDX biobanks aggregate these screens across many tumors and therapies~\citep{matschos:etal:2021:hroc-pdx-biobank,meric:etal:2024:pdxnet,akemi:etal:2025:pediatric-pdx-biobank}. Even so, exhaustive screening of every PDX-treatment combination remains infeasible due to cost, time, and labor constraints. Computational models that impute and rank the untested combinations can ease this burden by allowing researchers to analyze potential experiments \emph{in silico}. Such methods support \emph{drug repositioning}: the identification of new therapeutic uses for existing drugs.

\begin{figure}[t]
    \centering
    \includegraphics[width=\linewidth, trim=0 0.25cm 0 0.25cm,clip]{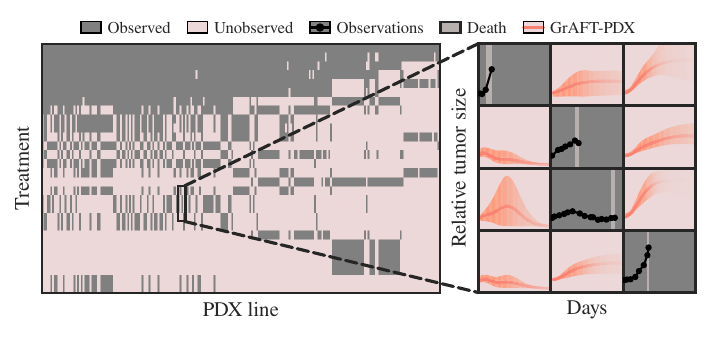}
    \caption{The left panel shows the sparsity of recorded experiments in the PDXe dataset relative to all possible PDX-treatment combinations. The right panel zooms in, showing how \modelname is capable of predicting responses for unobserved combinations from sparse, irregular measurements of previously conducted experiments.}
    \label{fig:intro-sparsity}
\end{figure}

We study the problem of computational modeling in the PDX Encyclopedia (PDXe)~\citep{Gao2015}, a large-scale PDX drug screen containing $223$ PDX lines, $62$ treatments, and $4190$ tested PDX-treatment combinations. \Cref{fig:intro-sparsity} illustrates both the incompleteness of the screen and the prediction task considered in this paper. Modeling the PDXe data poses three challenges: measurements are sparse both across and within combinations, tumor growth and survival are coupled through the observation process, and treatment response is heterogeneous across PDX lines. We consider each challenge in turn.

The sparsity of PDXe arises at two levels. First, many combinations are never tested; only \(4190\) of a possible \(13826\, (30.3\%)\) PDX-treatment pairs are tested. Second, even within tested combinations, tumor size is recorded on only \(29.8\%\) of days, on average, over a mouse's lifespan. Together, these leave the overall dataset \(91.7\%\) sparse. A model for these data must therefore impute at two levels, predicting both unobserved combinations and unmeasured points along each continuous tumor trajectory.

The coupling between growth and survival arises from the observation process itself. Tumor trajectories are only partially observed because measurements stop when a mouse dies. This missingness is informative. Mice with aggressive tumors tend to die earlier and therefore contribute no later measurements. Modeling tumor growth independently of survival would then bias estimated growth curves downward, since the fastest growing tumors are absent at later times. Capturing the dual nature of effective therapies---shrinking tumors and extending lifespan---requires joint modeling of longitudinal tumor growth and mouse survival.

Finally, treatment response is heterogeneous across PDX lines. The same therapy can shrink tumors in some lines yet appear far less effective in others, producing different trajectories and different lifespans. This variation is hard to predict from observed molecular covariates alone. This is true even in precision oncology, where genomic matching is an imperfect guide to response, with only around 63\% of patients responding to targeted therapies selected from tumor genomics~\citep{haslam:etal:2021:eligibility-response-targeted-therapy}. Capturing this heterogeneity therefore requires a model that learns PDX line specific response structure directly from the screening data rather than relying on observed covariates.

\subsection{Contributions}
Motivated by these challenges, we introduce \modelname, a Bayesian model that jointly predicts tumor growth and survival in sparse PDX drug screens. \modelname represents the sparse, irregular tumor measurements as noisy observations of an underlying smooth continuous-time trajectory and allows tumor growth to inform death risk. This couples the longitudinal and survival outcomes while accounting for the informative termination of tumor measurements at death. \modelname uses a low-rank tensor decomposition over PDX lines, treatments, and time to share information across observed experiments, predict untested PDX-treatment combinations, and capture heterogeneous treatment responses across PDX lines. The model represents tumor burden as a continuous trajectory and links its history to the survival process providing a novel description of how tumor burden impacts survival in mouse models. On PDXe, \modelname outperforms survival baselines at predicting time of death. For tumor response prediction, it is competitive with strong baselines for held-out experiments and can improve upon static baselines by incorporating partial tumor trajectory information. Although the model uses no genomic covariates, the predictions it generates preserve known associations between genes and drugs, suggesting that it learns biologically meaningful response structure from PDXe. We demonstrate \modelname can predict untested PDX-treatment combinations and rationally rank candidates for experimental follow-up.

\section{Related Work}
\label{sec:literature}
A large body of work in computational drug repositioning seeks to identify new therapeutic uses for existing compounds by analyzing molecular, phenotypic, or pharmacological data~\citep{jarada_review_2020}. Relational prediction methods have approached this problem by imputing unobserved drug-target or drug-disease associations, including probabilistic matrix completion~\citep{peska_drug-target_2017} and methods incorporating side information or causal structure~\citep{yang_drugdisease_2014,zhang_drimc_2020}. Closely related work in cancer predicts treatment response from molecular characteristics of tumors or cell lines~\citep{Mourragui2021}. On the PDXe dataset used here, \citet{Kurilov2020} compare prediction strategies that represent xenograft response through summaries such as tumor volume, tumor growth slope, and integrated response, while \citet{Gupta2025} reduce response to a binary outcome predicted by a single classifier across cancer types and treatments. These approaches predict treatment response for unobserved PDX-treatment combinations using scalar summaries of tumor response. This reduction discards the longitudinal structure of the tumor measurements, motivating models that instead represent the entire tumor trajectory over the course of treatment.

\modelname builds on the \emph{joint modeling} framework by linking time-varying tumor sizes to survival. Joint modeling is a statistical approach where longitudinal measurements and event times depend on a shared latent process \citep{ibrahim_basic_2010, Wang2025}. \citet{rizopoulos_dynamic_2017} model latent subject specific biomarker trajectories using linear mixed models and link them to a proportional hazards submodel. \citet{li_dynamic_2019} extend this framework to functional data by incorporating scalar and functional predictors into both the longitudinal and proportional hazards submodels, with penalized splines representing the functional effects. \citet{alvares_tractable_2021} combine a generalized linear mixed model with a parametric hazard survival model, linking the processes through shared fixed and random effects. In xenograft experiments, \citet{Pan2014} model longitudinal log tumor volume using a constrained linear mixed model and link its mouse specific random effect to a Cox proportional hazards model for survival. More recently, \citet{Naguib2026} link patient specific tumor trajectories to clinical event models to forecast overall survival from immature oncology trials. These approaches capture dependence between repeated measurements and event times, but do not naturally extend to sparse designs such as PDX screens, in which entire PDX-treatment combinations are unobserved.

We also utilize matrix and tensor factorization methods for predicting treatment responses. Large cancer drug screens form sparse response arrays indexed by biological samples and treatments, motivating constrained models that share information across these axes. Matrix factorization methods predict scalar drug response summaries across cancer cell lines and treatments by sharing information across observed combinations~\citep{AmmadUdDin2014,Suphavilai2018,Ren2022}. Adding experimental dimensions such as dose or drug combinations extends these matrix factorization approaches to tensor models. \citet{Wang2021} use latent tensor reconstruction to model responses across drugs, concentrations, and cell lines. \citet{Bertsimas2023} utilize tensor completion across patients, drugs, and doses to predict potential therapies for cancer patients. \citet{tansey:etal:2022:bayesian-dose-response} use Bayesian tensor filtering to model dose-response curves across cell lines, drugs, and doses. Together, these methods show how low-rank structure can be used to predict missing responses in sparsely observed cancer drug screens, including responses that vary across experimental dimensions such as dose and drug combinations. However, they do not model how treatment response evolves over time nor do they handle the informative censoring issue of in vivo studies. 

\section{PDXe Data and Notation}
\label{sec:data}
Each observed mouse $i$ is associated with a PDX line $\ell_i \in \{1,\ldots,L\}$ and a treatment $j_i \in \{0,1,\ldots,J\}$, where $j_i=0$ denotes the untreated control and values $j_i=1,\ldots,J$ denote active treatments or treatment combinations. Multiple mice may share the same PDX-treatment pair, though in PDXe only \(3\) out of the observed \(4190\, (0.072\%)\) pairs contain replicate mice.

Tumor measurements are taken at irregularly spaced times. For mouse $i$, we record both the raw tumor volume and its volume relative to the starting volume. We collect these quantities in the mouse level tumor record
\begin{equation}
    \mathcal{Y}_i = \left\{\ell_i, j_i, \bigl(r_i^{\mathsmaller{(m)}},y_i^{\mathsmaller{(m)}},\tau_i^{\mathsmaller{(m)}}\bigr)_{m=1}^{M_i}\right\}, \quad
    y_i^{\mathsmaller{(m)}} = \frac{r_i^{\mathsmaller{(m)}}}{r_i^{\mathsmaller{(1)}}}, \quad
    \tau_i^{\mathsmaller{(1)}} < \cdots < \tau_i^{\mathsmaller{(M_i)}}.
\label{eq:mouse-tumor-record}
\end{equation}
Here, $r_i^{\mathsmaller{(m)}}$ is the raw tumor volume measured at time $\tau_i^{\mathsmaller{(m)}}$, $y_i^{\mathsmaller{(m)}}$ is the corresponding relative tumor size, and $M_i$ is the number of tumor measurements for mouse $i$. Times are measured in days from treatment initiation, so $\tau_i^{\mathsmaller{(1)}}=0$ for all \(i\). The remaining measurement times may differ across mice. By construction, $y_i^{\mathsmaller{(1)}}=1$, and subsequent measurements describe multiplicative change from the initial observed tumor volume.

This normalization accounts for variability in initial tumor size. In PDXe, treatment began once tumors reached approximately $200\,\mathrm{mm}^{\mathsmaller{3}}$, but tumors were only monitored at most daily so volume is not identical across mice as some tumors may be larger when treatment begins. Modeling relative tumor size removes baseline tumor volume as a nuisance source of variation and places mice on a common response scale. This convention is also consistent with standard PDX efficacy analyses, where treatment response is often summarized through tumor growth inhibition~\citep{Wu2010-zh}, time-to-event endpoints based on tumor volume doubling~\citep{Saridogan2023}, or response categories defined relative to baseline tumor burden~\citep{Gao2015}.

Each mouse's observation sequence also encodes survival information. Let $t_i^{\mathsmaller{*}}$ denote the true latent death time for mouse $i$. In the PDXe data, every mouse is followed until death. The exact death time is not observed. Instead, $\tau_i^{\mathsmaller{(M_i-1)}}$ is the last time mouse $i$ was observed alive, and $\tau_i^{\mathsmaller{(M_i)}}$ is the first time mouse $i$ was observed to have died:
\begin{equation}
    \tau_i^{\mathsmaller{(M_i-1)}} < t_i^{\mathsmaller{*}} \leq \tau_i^{\mathsmaller{(M_i)}} .
\label{eq:death-interval}
\end{equation}
The relation above is the observed death interval for mouse $i$. Here $\tau_i^{\mathsmaller{(M_i)}}$ is also the final tumor measurement time, after which no further measurements are observed.

\section{Model}
\label{sec:model}
\modelname jointly models the tumor measurements and interval observed death time for each mouse. For mouse \(i\), the associated PDX line \(\ell_i\) and treatment \(j_i\) determine a continuous time median tumor trajectory and survival distribution. The observed measurements are noisy evaluations of this trajectory, and its previous values influence the mouse's death risk over time. We define the tumor and death submodels below.

\subsection{Tumor submodel}
\label{subsec:tumor_submodel}
The tumor submodel describes how relative tumor size evolves over time. Because relative tumor measurements are strictly positive and tumor growth is naturally multiplicative, we use a lognormal likelihood. Mouse \(i\)'s PDX line \(\ell_i\) and treatment \(j_i\) determine its tumor trajectory. We model the observed relative tumor measurements as
\begin{equation}
\label{eq:tumor-likelihood}
y_i^{\mathsmaller{(m)}} \sim \operatorname{Lognormal}\bigl(
\textsc{median}=G_{\ell_i,j_i}\bigl(\tau_i^{\mathsmaller{(m)}}\bigr),
\, \textsc{scale}=D\bigl(\tau_i^{\mathsmaller{(m)}}\bigr)
\bigr),
\quad m=2,\ldots,M_i.
\end{equation}
Here, \(G_{\ell_i,j_i}(t)\) is the median relative tumor size at time \(t\) for mice assigned to PDX line \(\ell_i\) and treatment \(j_i\), while \(D(t)\) controls the amount of measurement variation around this median. The likelihood begins at \(m=2\) because \(y_i^{\mathsmaller{(1)}}=1\) deterministically. The notation
\(X \sim \operatorname{Lognormal}\bigl(\textsc{median}=\mu,\textsc{scale}=d\bigr)\)
denotes that \(\log X \sim \normal\bigl(\log \mu,d^2\bigr)\). We parameterize the median and scale functions as smoothly evolving over time by modeling them with a fixed cubic B-spline basis~\citep{deboor1978},
\begin{equation}
\label{eq:tumor-functions}
G_{\ell,j}(t) \defeq \sum_{c=1}^C \theta_{\ell,j,c} B_c(t),
\qquad
D(t) \defeq \sum_{c=1}^C \eta_c B_c(t).
\end{equation}

We constrain each $\eta_c > 0$ to ensure that \(D(t)>0\). The coefficients \(\eta_1,\ldots,\eta_C\) determine how measurement variation changes over time and these coefficients have the priors, 
\begin{equation}
\eta_c \sim \mathrm{Exponential}(1).    
\end{equation}
We share \(D(t)\) across all PDX-treatment pairs because pair specific scale functions would be difficult to learn from the sparse observations.

\begin{figure}
    \centering
    \includegraphics[width=\linewidth, trim=0 0.25cm 0 0.25cm,clip]{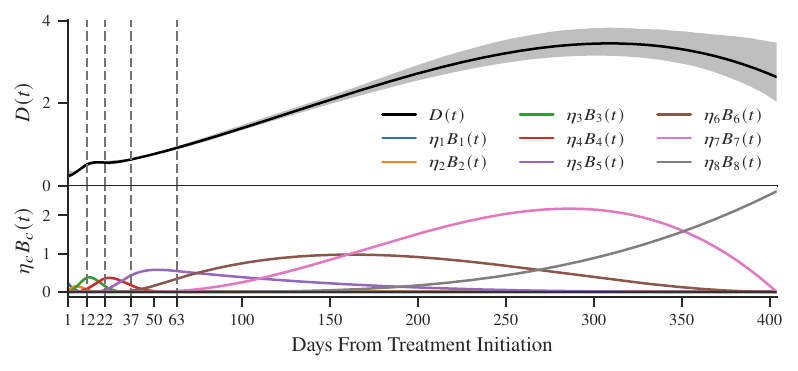}
    \caption{Sums of basis splines result in the learned time-varying scale. Interior knots denoted by dashed lines are placed at empirical quintiles of the observed measurement times. Each basis function is active only on a local interval, with neighboring basis functions overlapping.}
    \label{fig:spline_basis}
\end{figure}

We fix the basis functions before fitting the model; the model learns only their coefficients. A B-spline basis is determined by its degree and knot sequence. The degree controls the smoothness of the basis functions, while the knots determine where the basis functions are locally active. For a cubic B-spline basis with \(C\) basis functions, we write the knot sequence as \(\kappa_1\leq\cdots\leq\kappa_{C+4}\). The basis function \(B_c(t)\) is supported on \([\kappa_c,\kappa_{c+4}]\), so it is zero outside this interval and can affect a fitted function only over a local region of time. At any time \(t\), at most four cubic B-spline basis functions are nonzero. Neighboring basis functions overlap, allowing fitted functions to vary smoothly across time while retaining local control over different parts of the curve. \Cref{fig:spline_basis} illustrates how these basis functions come together to create $D(t)$.

Having specified the shared scale function \(D(t)\), we now turn to the median tumor trajectories \(G_{\ell,j}(t)\). We collect their spline coefficients in the tensor \(\pmb{\Theta} \in \mathbb{R}_+^{L\times(J+1)\times C}\), with entry \(\theta_{\ell,j,c}\) giving the coefficient on basis function \(B_c(t)\) for PDX line \(\ell\) under treatment \(j\). We write \(\pmb{\theta}_{\ell,j} = (\theta_{\ell,j,1},\ldots,\theta_{\ell,j,C}) \in \mathbb{R}_+^C\) for the corresponding fiber of \(\pmb{\Theta}\), which collects the coefficients defining \(G_{\ell,j}(t)\).

Rather than modeling \(\pmb{\Theta}\) as an unrestricted \(L\times(J+1)\times C\) tensor, \modelname represents it using a rank-\(K\) nonnegative canonical polyadic (CP) decomposition~\citep{Carroll1970,Harshman1970,Kolda2009}. A CP decomposition expresses a tensor as a sum of rank-one outer products, with each component coupling one loading pattern from each mode. Nonnegative loadings prevent components from canceling one another, so we can examine how each component contributes additively. Historically, nonnegative CP decompositions have been used both to predict unobserved entries in sparse relational tensors~\citep{Schein2015} and to recover interpretable biological patterns~\citep{Armingol2022}. These successful applications motivate using a nonnegative CP decomposition to predict unobserved responses and examine biological patterns in PDXe. Applied to \(\pmb{\Theta}\), each of the \(K\) components combines a pattern of loadings across PDX lines, a pattern across treatments, and a pattern across the spline basis. We write 
\begin{equation}
\label{eq:tumor-cp-factorization}
\theta_{\ell,j,c} \defeq \sum_{k=1}^K u_{\ell,k}v_{j,k}w_{c,k}, \qquad \pmb{\Theta} \defeq \sum_{k=1}^K \pmb{u}_k \circ \pmb{v}_k \circ \pmb{w}_k.
\end{equation}
We let \(\circ\) denote the outer product. Each term \(\pmb{u}_k \circ \pmb{v}_k \circ \pmb{w}_k\) is a rank-one tensor over PDX lines, treatments, and basis functions. The vector \(\pmb{u}_k\in\R^L\) contains the PDX line loadings for latent component \(k\), \(\pmb{v}_k\in\R^{J+1}\) contains the treatment loadings, and \(\pmb{w}_k\in\R^C\) contains the basis function loadings. The integer \(K\) is the number of rank-one components in the CP decomposition. When \(K\) is small relative to \(L\), \(J+1\), and \(C\), this representation replaces the unrestricted collection of \(L(J+1)C\) tumor trajectory coefficients with \(K(L+(J+1)+C)\) loading parameters. In practice, we use \(K=5\); see \Cref{subsec:choosingK} for details.

The CP decomposition also has a functional interpretation. Combining \cref{eq:tumor-functions,eq:tumor-cp-factorization} gives us a view on how \modelname is representing latent functions:
\begin{equation}
\label{eq:tumor-function-level-interpretation}
G_{\ell,j}(t)
=
\sum_{c=1}^C
\underbrace{
\biggl(
\sum_{k=1}^K
u_{\ell,k}v_{j,k}w_{c,k}
\biggr)
}_{=\,\theta_{\ell,j,c}}
B_c(t)
=
\sum_{k=1}^K
u_{\ell,k}v_{j,k}
\underbrace{
\biggl(
\sum_{c=1}^C
w_{c,k}B_c(t)
\biggr)
}_{\defeq\,\Phi_k(t)}.
\end{equation}
This shows that each latent factor \(k\) defines a shared tumor response function \(\Phi_k(t)\). The product \(u_{\ell,k}v_{j,k}\) weights the shared function \(\Phi_k(t)\) for pair \((\ell,j)\). Thus, \(G_{\ell,j}(t)\) is represented as a sum of shared latent functions whose weights depend on PDX line \(\ell\) and treatment \(j\).

We enforce positivity through nonnegative factor loadings. Specifically, we place independent exponential priors on the factor loadings,
\begin{equation}
u_{\ell,k}, v_{j,k}, w_{c,k} \sim \mathrm{Exponential}(1).    
\end{equation}
The exponential prior restricts the factor loadings in \cref{eq:tumor-cp-factorization} to be positive because the distribution only has support over \(\R_+\). Since cubic B-spline basis functions are nonnegative, the shared functions \(\Phi_k(t)\) in \cref{eq:tumor-function-level-interpretation} are also nonnegative. The products \(u_{\ell,k}v_{j,k}\) are positive as well, so the resulting median tumor trajectory \(G_{\ell,j}(t)\) is positive over the experimental time window. This is required because \(G_{\ell,j}(t)\) represents median relative tumor size, and relative tumor size is strictly positive. Importantly, positivity does not impose monotone tumor growth. The spline coefficients determine the height of local basis function contributions, not the slope of the trajectory. As the active basis functions change over time, \(G_{\ell,j}(t)\) can increase, decrease, or remain approximately stable while remaining positive.

\subsection{Death submodel}
\label{subsec:death_submodel}
The death submodel assigns probability to the interval of death for each mouse. Let \(t_i^{\mathsmaller{*}}\) denote the true latent death time for mouse \(i\). As described in \cref{eq:death-interval}, the observed survival information is \(\tau_i^{\mathsmaller{(M_i-1)}} < t_i^{\mathsmaller{*}} \leq \tau_i^{\mathsmaller{(M_i)}}\), where \(\tau_i^{\mathsmaller{(M_i-1)}}\) is the last time mouse \(i\) was observed alive and \(\tau_i^{\mathsmaller{(M_i)}}\) is the first time mouse \(i\) was observed to have died. Since mouse \(i\) is associated with PDX line \(\ell_i\) and treatment \(j_i\), its death distribution is determined by the survival function \(S_{\ell_i,j_i}(t)\), defined
\begin{equation}
    S_{\ell_i,j_i}(t)
    \defeq
    \Pr\bigl(t_i^{\mathsmaller{*}}>t \mid \ell_i,j_i\bigr).
\label{eq:mouse-survival-function}
\end{equation}
The survival function \(S_{\ell, j}(t)\) is a probability indexed by time. Therefore, for any given $t$ it must be the case that \(0 \leq S_{\ell,j}(t) \leq 1\). \(S_{\ell,j}(t)\) must also be monotonic decreasing in \(t\), as the event of remaining alive beyond a later time is contained in the event of remaining alive beyond any earlier time. Since all mice are alive at treatment initiation, we also have \(S_{\ell,j}(0)=1\). These constraints ensure that survival probabilities assigned to time intervals are valid. In particular, the probability that mouse \(i\) dies in its observed interval becomes
\begin{equation}
    \Pr\Bigl(
        \tau_i^{\mathsmaller{(M_i-1)}} < t_i^{\mathsmaller{*}}
        \leq \tau_i^{\mathsmaller{(M_i)}}
        \mid \ell_i,j_i
    \Bigr) 
    =S_{\ell_i,j_i}\bigl(\tau_i^{\mathsmaller{(M_i-1)}}\bigr)
    -
    S_{\ell_i,j_i}\bigl(\tau_i^{\mathsmaller{(M_i)}}\bigr).
\label{eq:death-interval-likelihood}
\end{equation}
Directly modeling \(S_{\ell,j}(t)\) with an unconstrained spline would not automatically enforce these properties: the fitted curve could exceed one, become negative, or increase locally, producing invalid interval probabilities in \cref{eq:death-interval-likelihood}. We therefore define the survival function indirectly through a nonnegative hazard function.

The hazard is a standard modeling object in survival analysis. Rather than modeling the probability of being alive beyond time \(t\) directly, the hazard describes the instantaneous death rate among mice that have survived up to time \(t\). For a fixed PDX line \(\ell\) and treatment \(j\), let \(\lambda_{\ell,j}(t)\) denote this hazard. For differentiable survival functions, this hazard can also be written as the instantaneous rate of decrease in log survival:
\begin{equation}
    \lambda_{\ell,j}(t)
    =
    \lim_{\Delta\downarrow 0}
    \frac{
        \Pr\bigl(t < z^{\mathsmaller{*}} \leq t+\Delta \mid z^{\mathsmaller{*}}>t,\ell,j\bigr)
    }{\Delta}
    =
    -
    \frac{d}{dt}
    \log S_{\ell,j}(t).
\label{eq:hazard-definition}
\end{equation}
Here \(z^{\mathsmaller{*}}\) denotes a generic latent death time for a mouse from PDX line \(\ell\) under treatment \(j\). Any nonnegative hazard defines an associated survival function
\begin{equation}
    S_{\ell,j}(t)
    =
    \exp\paren{
        -\int_0^t \lambda_{\ell,j}(u)\,du
    },
\label{eq:survival-from-hazard}
\end{equation}
which is automatically positive, bounded above by one, and nonincreasing. Thus, instead of directly modeling a highly constrained \(S_{\ell,j}(t)\), we model a nonnegative hazard and obtain a valid survival function through \cref{eq:survival-from-hazard}.

We model the hazard as the sum of two nonnegative components:
\begin{equation}
    \lambda_{\ell,j}(t)
    \defeq
    \lambda_{\ell,j}^{\mathrm{base}}(t)
    +
    \lambda_{\ell,j}^{\mathrm{tumor}}(t).
\label{eq:additive-hazard}
\end{equation}
The baseline component \(\lambda_{\ell,j}^{\mathrm{base}}(t)\) captures the underlying death risk for PDX line \(\ell\) under treatment \(j\), including survival patterns not explicitly attributed to the tumor trajectory. The tumor history component \(\lambda_{\ell,j}^{\mathrm{tumor}}(t)\) captures additional risk associated with the latent tumor trajectory up to time \(t\). Both components are constrained to be nonnegative, so each can only add to the instantaneous death rate in \cref{eq:additive-hazard}.

We choose this additive specification, following the linear hazard models of \citet{aalen1980nonparametric,aalen1989linear}. Whereas standard additive hazard models allow the current value of a time-varying covariate to contribute to death risk, our tumor history term depends on the latent tumor trajectory over all previous times. Under the spline representations introduced below, an additive hazard lets us precompute all required basis integrals and convolutions before inference. A proportional hazards formulation would introduce a nonlinear, nested time integral that cannot be computed before inference and would require numerical integration for each likelihood and gradient evaluation, making posterior inference computationally impractical.

The baseline hazard is represented using the same cubic B-spline basis introduced in \Cref{subsec:tumor_submodel}. For PDX line \(\ell\) and treatment \(j\), we define
\begin{equation}
    \lambda_{\ell,j}^{\mathrm{base}}(t)
    \defeq
    \sum_{c=1}^C
    \underbrace{\exp\paren{
    \sum_{k=1}^K
        u_{\ell,k}^{\mathsmaller{(d)}}
        v_{j,k}^{\mathsmaller{(d)}}
        w_{c,k}^{\mathsmaller{(d)}}
    }}_{\defeq\,\theta_{\ell,j,c}^{\mathsmaller{(d)}}}
    B_c(t).
\label{eq:baseline-hazard}
\end{equation}
The exponential transformation above ensures that each basis coefficient \(\theta_{\ell,j,c}^{\mathsmaller{(d)}}\) is positive. Since the B-spline basis functions are nonnegative, the resulting baseline hazard \(\lambda_{\ell,j}^{\mathrm{base}}(t)\) is nonnegative over the modeled time window. This lets the death factors \(u_{\ell,k}^{\mathsmaller{(d)}}, v_{j,k}^{\mathsmaller{(d)}}, w_{c,k}^{\mathsmaller{(d)}}\) vary freely on the real line, so we place standard normal priors on them, 
\begin{equation}
    u_{\ell,k}^{\mathsmaller{(d)}}, v_{j,k}^{\mathsmaller{(d)}}, w_{c,k}^{\mathsmaller{(d)}} \sim \normal\paren{0,1}.
\end{equation}

The tensor of log-coefficients \(\bigl(\log\theta_{\ell,j,c}^{\mathsmaller{(d)}}\bigr)\) is represented by a rank-\(K\) CP decomposition separate from the nonnegative CP decomposition of the tumor trajectory coefficient tensor \(\pmb{\Theta}\) in \cref{eq:tumor-cp-factorization}. Here \(u_{\ell,k}^{\mathsmaller{(d)}}\) is the loading of PDX line \(\ell\) on baseline hazard component \(k\), \(v_{j,k}^{\mathsmaller{(d)}}\) is the corresponding loading of treatment \(j\), and \(w_{c,k}^{\mathsmaller{(d)}}\) determines how that component contributes through basis function \(B_c(t)\). Thus, we use a nonnegative CP decomposition for the tumor coefficient tensor and an unconstrained CP decomposition for the baseline hazard log-coefficient tensor, then exponentiate the latter elementwise. \Cref{sec:parameter-constraints} compares this construction with alternatives that use the same positivity mechanism in both submodels.

The tumor history component links the death process to the latent tumor trajectory. While \(\lambda_{\ell,j}^{\mathrm{base}}(t)\) captures baseline death risk, \(\lambda_{\ell,j}^{\mathrm{tumor}}(t)\) captures additional risk associated with the latent tumor trajectory up to time \(t\). Inspired by the weighted cumulative exposure model of \citet{Sylvestre2009}, we represent this dependence using a convolution term in the hazard. For PDX line \(\ell\) under treatment \(j\), we define
\begin{equation}
    \lambda_{\ell,j}^{\mathrm{tumor}}(t)
    \defeq
    \int_0^t
        \beta_{\mathrm{conv}}\bigl(t-s\bigr)
        G_{\ell,j}(s)
    \,ds,
    \qquad
    \beta_{\mathrm{conv}}(a)
    \defeq
    \sum_{c=1}^C
        \beta_c B_c(a).
\label{eq:tumor-history-hazard}
\end{equation}
Here, \(G_{\ell,j}(s)\) is the latent median relative tumor size at previous time \(s\), and \(\beta_{\mathrm{conv}}\) is a shared lag response kernel represented in the same cubic B-spline basis as $G_{\ell,j}$. The argument \(t-s\) is the lookback time from the current time \(t\) to the tumor burden at time \(s\). Equivalently, writing \(a=t-s\), larger values of \(\beta_{\mathrm{conv}}(a)\) indicate lags at which historical tumor burden contributes more strongly to current death risk. Thus, \(\lambda_{\ell,j}^{\mathrm{tumor}}(t)\) is a weighted summary of previous tumor burden, where the shared kernel \(\beta_{\mathrm{conv}}\) determines which parts of the past trajectory are most relevant for current death risk. Importantly, we model the full tumor burden history because longitudinal measurements can provide survival information that is lost when tumor response is reduced to a scalar summary \citep{Zhou2023}. 

Expanding out $\beta_{\textrm{conv}}(a)$ and $G_{\ell,j}(t)$ in \cref{eq:tumor-history-hazard} gives
\begin{equation}
    \lambda_{\ell,j}^{\mathrm{tumor}}(t)
    =
    \sum_{c = 1}^C\sum_{d=1}^C
        \beta_c\theta_{\ell,j,d}
        \int_0^t
            B_c(t-s)B_d(s)
        \,ds.
\label{eq:tumor-history-basis-convolution}
\end{equation}
Although \(\lambda_{\ell,j}^{\mathrm{tumor}}(t)\) is written as a double sum over all basis function pairs, most terms vanish at a given time \(t\). The factor \(B_d(s)\) is nonzero only on its local support \(s \in [\kappa_d, \kappa_{d+4}]\), while \(B_c(t-s)\) is nonzero only when \((t-s) \in [\kappa_c, \kappa_{c+4}]\). Thus, the integrand is nonzero only on the portions of \([0,t]\) where these two local supports overlap. On each overlap, both basis functions have a polynomial form, so when nonzero \(B_c(t-s)B_d(s)\) is a polynomial in \(s\). Consequently, each integral in \cref{eq:tumor-history-basis-convolution} reduces to a finite sum of piecewise polynomial integrals on regions of overlap.

We place an index biased shrinkage prior on each kernel coefficient,
\begin{equation}
    \beta_c
    \sim
    \mathrm{Exponential}\left(
        10^{c-1}
        \int_0^T B_c(a)\,da
    \right).
\label{eq:kernel-prior}
\end{equation}
Under the standard parameterization of the exponential distribution, \(\mathrm{Exponential}(\lambda)\) has mean \(1/\lambda\). Therefore \cref{eq:kernel-prior} implies
\begin{equation}
    \E{\beta_c}{
        \int_0^T
            \beta_c B_c(a)
        \,da
    }
    =
    10^{1-c}.
\label{eq:kernel-prior-component-mass}
\end{equation}
Thus, the factor \(10^{1-c}\) sets the prior expected total weight of the \(c\)-th kernel component. The normalization by \(\int_0^T B_c(a)\,da\) makes the prior act on integrated kernel weight rather than on raw coefficient magnitude. Since interior knots are placed at empirical quintiles of the observed measurement times, basis functions differ in support width, magnitude, and total area as shown in \Cref{fig:spline_basis}. Without normalization, basis functions with larger area would receive larger prior expected contribution to \(\beta_{\mathrm{conv}}\) solely because of the geometry of the chosen basis.

To understand the induced prior on the full convolution kernel, consider the kernel weight accumulated up to time \(t\).
\begin{equation}
    \E{\beta_1,\ldots,\beta_C}{
        \int_0^t
            \beta_{\mathrm{conv}}(t-s)
        \,ds
    }
    =
    \sum_{c=1}^C
        10^{1-c}
        \underbrace{
        \frac{
            \int_0^t B_c(a)\,da
        }{
            \int_0^T B_c(a)\,da
        }}_{\defeq F_c(t) \in [0,1]}.
\label{eq:kernel-prior-cumulative}
\end{equation}
Here \(F_c(t)\) is the fraction of the \(c\)-th basis function's area accumulated by time \(t\). Because the B-spline basis functions are local and ordered from recent to more distant lags, only a few nearby basis functions contribute to \(\beta_{\mathrm{conv}}(a)\) at each lag \(a\). At later lags, basis functions with larger indices \(c\) contribute most, and the corresponding kernel components have geometrically smaller expected total weights. The prior consequently imposes stronger shrinkage on distant tumor history effects, which are weakly informed because fewer mice survive long enough to contribute information about long lookback windows. The coefficients \(\beta_c\) are constrained to be nonnegative, so previous tumor burden can only increase the hazard; because these coefficients are shared across all PDX lines and treatments, \(\beta_{\mathrm{conv}}(a)\) provides a common summary of how tumor history contributes to survival.

\section{Inference and Synthetic Testing}
\label{sec:infernece}
The \modelname model induces a posterior distribution over the latent tumor trajectories, survival functions, measurement error function, and tumor history kernel. We generate posterior samples from this model using Hamiltonian Monte Carlo (HMC), and use those posterior draws to compute summary statistics.

\subsection{Posterior target and inference}
\label{sec:posterior-target}

Let $\Psi$ collect all unknown model parameters: the tumor factor loadings, death factor loadings, measurement scale coefficients, and tumor history kernel coefficients. That is,
\begin{equation}
\Psi \defeq \left\{u_{\ell,k}, v_{j,k}, w_{c,k},u^{\mathsmaller{(d)}}_{\ell,k}, v^{\mathsmaller{(d)}}_{j,k}, w^{\mathsmaller{(d)}}_{c,k},\eta_c, \beta_c\right\}.
\end{equation}
Given the mouse observation records defined in \cref{eq:mouse-tumor-record} where $\mathcal{Y}_i$ contains the relative tumor sizes and observation times for mouse $i$, posterior inference then targets
\begin{equation}
p\paren{\Psi \mid \mathcal{Y}_{1:N}} \propto
p\paren{\mathcal{Y}_{1:N}\mid\Psi} p(\Psi).
\label{eq:posterior-target}
\end{equation}
The likelihood term combines the lognormal tumor likelihood with the interval censored survival likelihood defined in \Cref{sec:model}. The prior $p(\Psi)$ is the product prior over all parameters.

Although the posterior is parameterized in terms of spline coefficients and low-rank tensor factors, these parameters are not the main inferential targets. Each posterior draw of $\Psi$ induces tumor quantities such as the median latent tumor trajectory $G_{\ell,j}(t)$ and scale function $D(t)$, survival quantities such as the hazard components $\lambda_{\ell,j}^{\mathrm{base}}(t)$ and $\lambda_{\ell,j}^{\mathrm{tumor}}(t)$ and the survival function $S_{\ell,j}(t)$, and the shared convolution kernel $\beta_{\mathrm{conv}}(a)$. For each posterior draw, we evaluate these functions, then summarize their values across draws using means or quantiles. For example, if $\Psi^{\mathsmaller{(1)}},\ldots,\Psi^{\mathsmaller{(R)}}$ are posterior draws, the posterior mean of the survival function for PDX line $\ell$ under treatment $j$ is estimated by
\begin{equation}
\widehat S_{\ell,j}(t)
\defeq
\frac{1}{R}
\sum_{r=1}^R
S_{\ell,j}\!\left(t;\Psi^{\mathsmaller{(r)}}\right)
\approx
\E{\Psi \sim p\paren{\Psi \mid \mathcal{Y}}}{
S_{\ell,j}(t;\Psi) \mid \mathcal{Y}}.
\label{eq:posterior-mean-survival}
\end{equation}
We use analogous Monte Carlo summaries for tumor trajectories, hazard functions, posterior predictive distributions, and the convolution kernel.

We implement \modelname in Stan~\citep{Stan} and fit the model using the No-U-Turn Sampler (NUTS), an adaptive HMC algorithm~\citep{Brooks_2011,hoffman14a}. For each fitted model, we run four chains with 1000 warmup iterations and 500 post-warmup sampling iterations per chain, yielding 2000 retained posterior draws. We use no thinning and set \texttt{max\_treedepth} to 7.

\subsection{Synthetic recovery of the tumor history kernel}
\label{subsec:synthetic_beta_recovery}

The convolution kernel $\beta_{\mathrm{conv}}(a)$ summarizes how previous tumor burden contributes to current death risk. Before applying \modelname to PDXe, we test whether it can recover a known kernel from synthetic data. This ensures that the model's structure does not introduce spurious relationships between tumor history and death and that the model can identify a relationship when one is present.

We generate synthetic datasets using the same time scale, spline basis, and knot locations as in our PDXe analysis in \cref{sec:results}. For each synthetic dataset, we first draw one set of tumor trajectory factors and baseline hazard factors from the model prior. This prior draw defines the latent tumor trajectories \(G_{\ell,j}^{\mathrm{true}}(t)\) and synthetic baseline hazards \(R_{\ell,j}^{\mathrm{true}}(t)\) for all observations. The synthetic death hazard is
\begin{equation}
\lambda_{\ell,j}^{\mathrm{true}}(t)
=
R_{\ell,j}^{\mathrm{true}}(t)
+
\int_0^t
\beta_{\mathrm{conv}}^{\mathrm{true}}(t-s)
G_{\ell,j}^{\mathrm{true}}(s)
\,ds.
\label{eq:synthetic-death-hazard}
\end{equation}
For each mouse, we draw a latent death time from the survival distribution induced by \cref{eq:synthetic-death-hazard}. Synthetic tumor measurements are generated from the lognormal observation model every three days. We use a three day spacing because it approximately matches the observation frequency in the PDXe data. Let \(t_i^{\mathsmaller{*}}\) be the exact synthetic death time and let \(\mathcal T=\{0,3,6,\ldots\}\) denote the observation grid. We record death through the interval \((a_i,b_i]\), where \(a_i=\max\{t\in\mathcal T:t<t_i^{\mathsmaller{*}}\}\) and \(b_i=\min\{t\in\mathcal T:t\ge t_i^{\mathsmaller{*}}\}\).

\begin{figure}[ht]
    \centering
    \includegraphics[width=\linewidth, trim=0 0.3cm 0 0.25cm,clip]{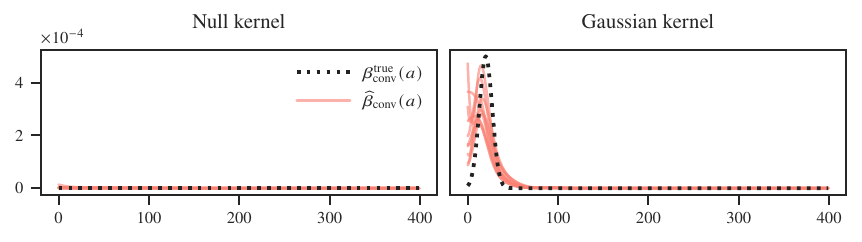}
    \caption{Across 10 independently generated synthetic datasets per kernel, \modelname recovers the localized tumor history effect in the Gaussian setting while returning estimates near zero under the null.}
    \label{fig:beta_recovery_metrics}
\end{figure}

We consider two data generating settings for \(\beta_{\mathrm{conv}}^{\mathrm{true}}(a)\), a null kernel and a Gaussian kernel,
\begin{equation}
\beta_{\mathrm{conv}}^{\mathrm{true}}(a)
\in
\left\{
\beta_{\mathrm{conv}}^{\mathrm{Null}}(a)=0,\quad
\beta_{\mathrm{conv}}^{\mathrm{Gauss}}(a)=5 \times 10^{-4}\exp\paren{-\frac{\paren{a-20}^2}{10^2}}
\right\}.
\label{eq:synthetic_kernel_settings}
\end{equation}
The two settings for $\beta_{\mathrm{conv}}^{\mathrm{true}}(a)$ target complementary recovery questions. The null kernel \(\beta_{\mathrm{conv}}^{\mathrm{Null}}(a)\) tests whether the fitted model avoids inferring a tumor history effect when no such effect is present. The Gaussian kernel \(\beta_{\mathrm{conv}}^{\mathrm{Gauss}}(a)\) has peak height \(5\times 10^{-4}\) at lag \(a=20\) days. Writing its exponent in the usual Gaussian form gives standard deviation \(10/\sqrt{2}\approx 7.1\) days, so approximately \(95\%\) of its mass is concentrated between lags 6 and 34 days. Thus, the Gaussian setting creates a localized effect centered about three weeks before the current hazard time and tests whether \modelname can recover both the existence of a tumor history effect and the region where that effect is concentrated. \Cref{fig:beta_recovery_metrics} summarizes these recovery tests. Visually, we find that \modelname consistently recovers the true effect of historical tumor burden on survival.

Quantitatively, we assess how accurately GrAFT-PDX recovers the cumulative tumor effect in the Gaussian case. Let
\begin{equation}
M=\int_0^T \beta^{\mathrm{Gauss}}_{\mathrm{conv}}(a)\,da
\qquad\text{and}\qquad
\widehat{M}=\int_0^T \widehat{\beta}_{\mathrm{conv}}(a)\,da
\end{equation}
denote the areas under the true and estimated kernels, respectively. We measure recovery of the cumulative effect using the ratio \(\widehat{M}/M\), for which a value of one indicates that the estimated and true kernels have equal total area. Across the ten datasets, \(\widehat{M}/M=0.97\) with a standard deviation of \(0.14\), indicating that the model recovered the true cumulative effect.

\section{Results}
\label{sec:results}
We evaluate \modelname on the PDXe dataset of \citet{Gao2015}. First, we assess held-out prediction of survival and tumor response under partially observed tumor trajectories, comparing \modelname to survival baselines, tumor response baselines, and simpler versions of the \modelname model. We then interrogate the learned convolution kernel, $\beta_{\mathrm{conv}}$, to assess how previous tumor burden contributes to current death risk. To evaluate the scientific usefulness of \modelname, we examine whether posterior predictive outcomes preserve drug-gene response patterns and whether \modelname suggests biologically coherent candidates for follow-up experiments.

\subsection{Held-out prediction on PDXe}

\begin{figure}[ht]
    \centering
    \includegraphics[width=\linewidth, trim=0 0.3cm 0 0.25cm,clip]{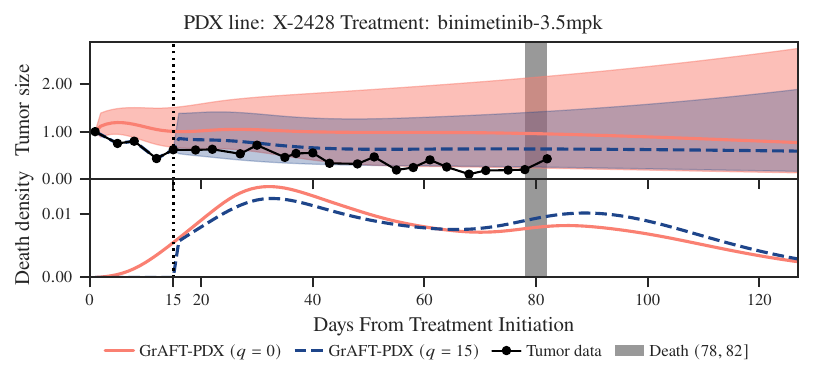}
    \caption{Conditioning on the first 15 days of tumor measurements shifts the posterior tumor trajectory toward the observed early response and removes posterior predictive death mass before day 15. Ribbons denote the 50\% credible interval. For tumor size, the predictions are posterior median whereas for death density they are the posterior predictive distribution.}
    \label{fig:prediction-censoring-example}
\end{figure}

We evaluate prediction using 20 train-test splits constructed at the PDX-treatment combination level. Within each split, we exclude all mice from each test combination from training. We evaluate predictions for each test mouse at censoring times \(q\in\{0,5,10,15\}\). The censoring times chosen are intentionally small as one use for \modelname is to identify unsuccessful PDX experiments early. At \(q=0\), the fitted model has observed no outcomes from the test combination. At \(q>0\), prediction conditions on the mouse's tumor measurements at times \(\tau_i^{\mathsmaller{(m)}} \leq q\) and on survival through \(q\). Its subsequent tumor measurements and death interval remain held out for evaluation. \Cref{fig:prediction-censoring-example} illustrates the information update induced by censoring. At \(q=0\), prediction is driven solely by the learned parameters, whereas at \(q=15\), the prediction additionally reflects the mouse's early observed tumor response. 

For mouse \(i\), implanted with PDX line \(\ell_i\) and assigned treatment \(j_i\), let \(S_i(t) \defeq S_{\ell_i,j_i}(t)\). Survival through the censoring time contributes \(S_i(q)\) to the conditioning likelihood by \cref{eq:mouse-survival-function}. Increasing \(q\) therefore provides additional early tumor measurements while shortening the unobserved portion of the mouse's trajectory.

For survival prediction, conditioning on survival through \(q\) truncates and renormalizes the posterior predictive death distribution. Let \(a_i=\tau_i^{\mathsmaller{(M_i-1)}}\) and \(b_i=\tau_i^{\mathsmaller{(M_i)}}\), so that the observed death interval is \((a_i,b_i]\). Conditional on a posterior draw of \(S_i\) and on survival through \(q\), the predictive survival function is
\begin{equation}
    \Pr\paren{t_i^{\mathsmaller{*}} > t \mid t_i^{\mathsmaller{*}} > q, S_i}
    =
    \frac{S_i(t)}{S_i(q)},
    \qquad t \geq q .
    \label{eq:conditional-survival}
\end{equation}
If the observed death interval for mouse \(i\) satisfies \(b_i>q\), then the conditional probability assigned to the future death interval is
\begin{equation}
    \Pr\paren{
        a_i < t_i^{\mathsmaller{*}} \leq b_i
        \mid
        t_i^{\mathsmaller{*}} > q, S_i
    }
    =
    \frac{
        S_i\paren{\max\{a_i,q\}}
        -
        S_i\paren{b_i}
    }{
        S_i\paren{q}
    } .
    \label{eq:conditional-death-interval}
\end{equation}
Mice with \(b_i \leq q\) are not included in the prediction task because their recorded death interval ends before or at the censoring time. In evaluation, we average the conditional interval probabilities across posterior draws to obtain posterior predictive probabilities.

We evaluate predictive performance using four complementary metrics. The death log probability evaluates the probability assigned to the observed future death interval after conditioning on survival through \(q\). Given a death interval \((a_i,b_i]\), this score is
\(
    \log
    \Pr\paren{
        a_i < t_i^\mathsmaller{*} \leq b_i
        \mid
        t_i^\mathsmaller{*} > q
    } .
\)
The log probability rewards models that assign high probability to the observed death interval. The median death mean absolute error (MAE) compares the posterior median predicted death time to the first recorded death time \(b_i\), giving an error in days. The death concordance index (C-index) evaluates ranking accuracy by measuring whether mice predicted to die earlier do in fact die earlier. For tumor response, we evaluate mRECIST classification accuracy using the clinical categories: complete response (CR), partial response (PR), stable disease (SD), and progressive disease (PD). This metric evaluates whether the model predicts clinically interpretable tumor response categories.

We compare \modelname to separate survival and tumor response baselines. The survival baselines are the Cox proportional hazards model \citep{Cox1972} equipped with the Breslow baseline hazard estimator \citep{Breslow72} and the Royston-Parmar spline-based survival model \citep{Royston2002-we}. The tumor response baselines are multinomial regression, a multilayer perceptron classifier, and XGBoost \citep{ChenG16}. Baseline hyperparameter values and tuning grids are given in \Cref{ref:HyperparamTuning}.

\begin{figure}[ht]
    \centering
    \includegraphics[width=\linewidth, trim=0 0.25cm 0 0.1cm,clip]{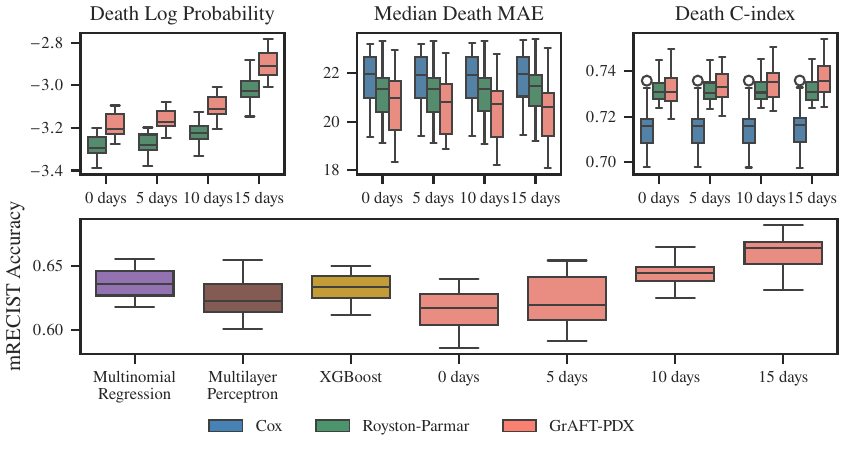}
    \caption{Predictions are evaluated at censoring times \(q \in \{0, 5, 10, 15\}\). The Cox model is omitted from the log-probability panel because the Breslow estimator for the baseline hazard assigns 0 probability to some held-out death intervals. For mRECIST accuracy, the \modelname results show how partial tumor trajectory information improves predictive performance. For survival, we see that \modelname outperforms baselines across all metrics with an increasing performance gap with increasing \(q\).}
    \label{fig:censoring-baselines}
\end{figure}

\Cref{fig:censoring-baselines} shows that \modelname achieves higher death log probabilities and lower death time prediction errors than the survival baselines across all censoring times. Relative to Royston-Parmar, its death log probability is higher by \(0.11\)--\(0.12\) nats, the MAE of its predicted median death time is lower by \(0.43\)--\(0.89\) days, and its C-index matches or exceeds that of Royston-Parmar by up to \(0.004\) across \(q\in\{0,5,10,15\}\). For tumor response prediction, the mean mRECIST accuracy of \modelname is \(61.6\%\) at \(q=0\), compared with \(63.6\%\) for the best static classification baseline, multinomial regression. From \(q\geq10\) it outperforms multinomial regression with a mean accuracy of \(64.5\%\) at \(q=10\) increasing to \(66.1\%\) at \(q=15\). As the mRECIST baselines cannot condition on partial tumor trajectories, the \(q>0\) results highlight a unique capability of \modelname among the evaluated models to update tumor predictions using the partial trajectory observed before censoring.

\subsection{Ablation study}
We next evaluate whether the tumor history component improves survival prediction on held-out data. We compare the full model to an ablated model that keeps the same baseline hazard model, spline basis, low-rank factorization, priors, train-test splits, and HMC settings, but removes the convolution term from the hazard. Under this ablation, \(\lambda_{\ell,j}(t) = \lambda_{\ell,j}^{\mathrm{base}}(t)\), so survival risk can vary across PDX lines, treatments, and time, but it cannot depend directly on the latent tumor trajectory. This particular ablation tests if the learned convolution kernel \(\beta_{\mathrm{conv}}(a)\) is necessary to make accurate predictions. Moreover, in order to use \modelname to truncate unsuccessful PDX experiments early, we must confirm that \(\beta_{\mathrm{conv}}(a)\) is the mechanism through which tumor size informs death risk. 

For each censoring time \(q \in \{0,5,10,15\}\), we evaluate both models using five-fold cross-validation under the held-out prediction task defined above. We partition the observed PDX-treatment combinations into five disjoint folds, assigning all mice from the same combination to a single fold. Each model is fit to four folds and evaluated on all mice belonging to combinations in the remaining fold, so every mouse contributes one out-of-sample prediction. For each test mouse \(i\), the death contribution is the log pointwise predictive density (LPPD) of its future death interval conditional on survival through \(q\), as defined in \cref{eq:conditional-death-interval}. The tumor contribution is the log joint posterior predictive density of all tumor measurements observed after \(q\). To isolate the predictive contribution of tumor history, we define the per mouse differences
\begin{equation}
    \Delta \mathrm{LPPD}_{i}^{o}(q)
    \defeq
    \mathrm{LPPD}_{i,\mathrm{full}}^{o}(q)
    -
    \mathrm{LPPD}_{i,\mathrm{no\ history}}^{o}(q),
    \qquad
    o \in \{\mathrm{death},\mathrm{tumor}\}.
    \label{eq:lppd-difference}
\end{equation}
The death difference measures whether the tumor history component improves held-out survival prediction, while the tumor difference checks whether including this component changes the quality of the fitted tumor trajectories. We summarize these differences by their mean and standard error across held-out mice.

\begin{figure}[ht]
    \centering
    \includegraphics[width=\linewidth, trim=0 0.25cm 0 0.25cm,clip]{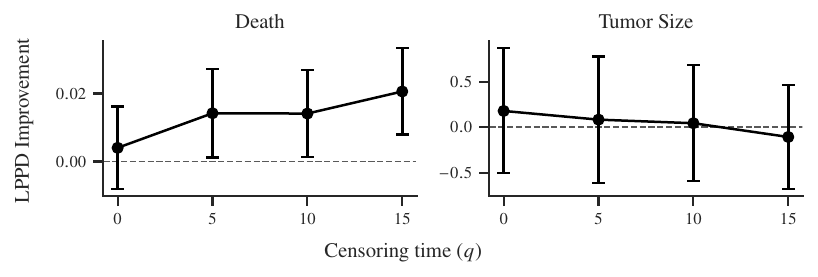}
    \caption{Per mouse differences in held-out predictive accuracy between the full model and the ablation at censoring times \(q \in \{0,5,10,15\}\). Positive values favor the full model and error bars represent the \(95\%\) confidence interval for the mean per mouse difference over a five-fold cross validation. Once early tumor measurements are available \((q \geq 5)\), the tumor history component improves death interval prediction without materially changing prediction of subsequent tumor measurements.}
    \label{fig:delta_lppd}
\end{figure}
\Cref{fig:delta_lppd} summarizes the mouse level differences in death and tumor LPPD. At \(q=0\), the mean death \(\Delta\mathrm{LPPD}\) is positive, but its \(95\%\) confidence interval includes zero. At \(q\in\{5,10,15\}\), the mean death \(\Delta\mathrm{LPPD}\) is positive and each \(95\%\) confidence interval excludes zero, indicating that the full model assigns greater probability to held-out death intervals than the ablated model once early tumor measurements are available. By contrast, the \(95\%\) confidence intervals for the mean tumor \(\Delta\mathrm{LPPD}\) include zero at every censoring time, providing no evidence of a systematic difference between the two models. Together with the synthetic recovery experiment, these results provide strong evidence that tumor history contributes to death risk in the PDXe data and that \modelname can detect and characterize such effects when they are present.

\subsection{Interpreting the learned tumor history kernel} The kernel \(\beta_{\mathrm{conv}}(a)\) weights the latent median tumor size \(G_{\ell,j}(t-a)\) when computing the hazard at time \(t\). Its shape describes how the weight assigned to a given tumor burden varies with the time elapsed since that burden occurred. Applied to PDXe, \(\beta_{\mathrm{conv}}(a)\) provides new insight into how important historical tumor burden is to current death risk. This construction follows weighted cumulative exposure and distributed lag models, which allow past covariate values to contribute differently according to their lag \citep{Sylvestre2009,Gasparrini2010}. Related joint models use weighted biomarker histories to link longitudinal measurements to survival \citep{Mauff2017}. We estimate the kernel from PDXe to identify which periods of tumor history are most strongly associated with current death risk.

\begin{figure}[ht]
    \centering
    \includegraphics[width=\linewidth, trim=0 0.25cm 0 0.2cm,clip]{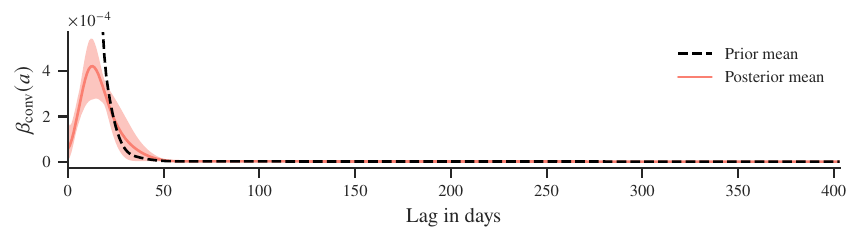}
    \caption{Posterior estimate of the tumor history kernel \(\beta_{\mathrm{conv}}(a)\), where \(a\) denotes the lag in days. Larger values assign greater weight to tumor burden observed at the corresponding lag when constructing the current tumor history hazard. The fitted kernel is concentrated at recent lags, with the posterior mean attaining its maximum at approximately two weeks. Crucially, we find that it does not decrease monotonically as we had initially expected.}
    \label{fig:TumorRelationship}
\end{figure}

As shown in \Cref{fig:TumorRelationship}, the fitted kernel places most of its mass on recent tumor history. Its posterior mean is small at lag zero, rises to its maximum at a lag of \(11\) days, and then declines at longer lags. Thus, the fitted model associates current death risk more strongly with tumor burden accumulated during the preceding weeks than with either instantaneous tumor size alone or more remote tumor history. Because the location of the maximum is inferred from a smooth, regularized kernel, we interpret the figure primarily as evidence for a short recent lag window rather than any specific delay.

This association is broadly consistent with earlier analyses, although those analyses impose different association structures. In xenograft experiments, \citet{Pan2014} linked longitudinal tumor volume and survival through shared random effects and found a positive association between tumor volume and death risk. In clinical oncology, \citet{Claret2009,Claret2013} identified early change in tumor size and time to tumor growth as predictors of overall survival, while \citet{Tardivon2019} found that both time to tumor growth and the instantaneous rate of tumor size change were informative. 

The posterior mean's peak away from zero indicates that elevated tumor burden may increase death risk through physiological deterioration that develops over time. Mouse experiments have documented metabolic disruption, muscle loss, and weight loss emerging over successive days \citep{OConnell2021}. These observations provide a plausible explanation for why earlier tumor burden could be associated with current death risk, although they do not establish which processes underlie the fitted kernel. Because PDXe does not measure all of these changes, evaluating this explanation would require further studies and analyses that track them alongside tumor volume and survival.

\subsection{Qualitative analysis of latent components}

\begin{figure}[ht]
\centering
\includegraphics[width=\textwidth, trim=0 0.3cm 0 0.25cm,clip]{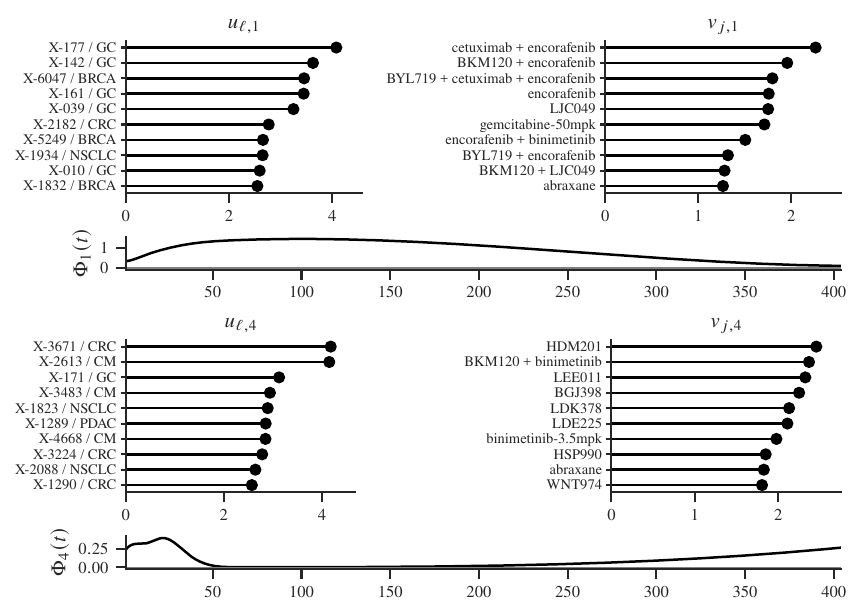}
\caption{Latent tumor components from a single draw of the \(K=5\) model. For each component, the panels display the ten PDX lines with the largest loadings, the ten treatments with the largest loadings, and its latent function.}
\label{fig:select-tumor-components}
\end{figure}

Nonnegative CP decompositions give us the ability to look inside of our model and examine its inferential components to ensure they admit plausible interpretations. In a good model, such qualitative analysis should reveal the biological mechanisms that underpin the statistical inference. We examine the \(K=5\) decomposition from the final posterior draw and visualize selected components in \cref{fig:select-tumor-components}. For completeness, \cref{subsec:AllTumorLatents} visualizes all latent components.

Among the five latent components, component 4 exhibits the clearest correspondence between the treatments and PDX lines. Two of its highly weighted treatments contain binimetinib, an inhibitor of MEK1 and MEK2, while three of its highly weighted PDX lines---X-2613, X-3483, and X-4668---are melanomas harboring the activating BRAF V600E mutation. Because BRAF signals through MEK in the MAPK pathway, this grouping agrees with experimental evidence that BRAF-mutant tumors are particularly sensitive to MEK inhibition \citep{solit2006braf}. The component also includes LEE011, an inhibitor of CDK4 and CDK6, together with four PDX lines---X-3671, X-2613, X-3483, and X-4668---having deletions of CDKN2A, CDKN2B, or both. These genes encode negative regulators of CDK4/6 activity, and loss of CDKN2A has previously been associated with sensitivity to CDK4/6 inhibition in melanoma \citep{Young2014}. The latent function decreases over the first 60 days, consistent with a shrinking tumor in a group containing established genomic relationships.

Component 1 provides a complementary example in which the treatments are negatively related to the top PDX loadings. Six of its ten highly weighted treatments contain encorafenib, a selective RAF inhibitor whose activity is concentrated in tumors harboring activating BRAF V600 mutations \citep{Delord2017}. Among the ten highly weighted PDX lines, however, only X-2182 carries BRAF V600E. Consistent with this genomic mismatch, the temporal factor increases early and remains substantial rather than shrinking. Component 1 may therefore represent sustained tumor growth under BRAF-directed treatment in a collection of PDX lines that largely lacks the relevant genomic mutation.

\subsection{Biological association preservation}

We next ask whether posterior predictive outcomes from \modelname retain genomic response patterns observed in PDXe. The model is fit using tumor trajectories and survival outcomes alone; genomic covariates are not included. We opt not to use genomic information as there is no clear method to incorporate it and the quantity of genomic information would make posterior inference computationally intractable. Therefore, we can use genomic data after fitting as an external check on the model-generated outcomes. To evaluate this, we compare association analyses fit to the observed data with the same analyses fit to a model-generated dataset.

The generated dataset is constructed out of sample. First, we partition the treated PDX-treatment pairs into five folds. For each fold, we train \modelname on the other four folds and use it to generate posterior predictive outcomes for the held-out treated pairs. We retain all untreated mice in every training set. Aggregating the held-out predictions across folds gives one generated dataset whose treated outcomes are predicted out of sample, though with weak dependence due to being constructed from the jackknife-style cross-validation procedure.

We analyze the observed and generated datasets in the same way. We first define the treatment effect using lifespan as the response. Let \(\bar{s}_{\ell,j}\) denote the average lifespan among mice from PDX line \(\ell\) assigned treatment \(j\), and let \(\bar{s}_{\ell,0}\) denote the average lifespan among untreated mice from the same PDX line. We define
\begin{equation}
    \Delta_{\ell,j}
    \defeq
    \bar{s}_{\ell,j}
    -
    \bar{s}_{\ell,0}.
    \label{eq:treatment-effect}
\end{equation}
Thus, \(\Delta_{\ell,j}\) compares lifespan under treatment \(j\) to no treatment for the same PDX line.

To relate the treatment effect to tumor genomic information, we fix a treatment \(j\), a gene \(g\), and a gene alteration type \(m \in \{\mathrm{amplification}, \mathrm{deletion}, \mathrm{mutation}\}\). Let \(x_{\ell,g}^{\mathsmaller{(m)}} \in \{0,1\}\) indicate whether PDX line \(\ell\) has alteration type \(m\) in gene \(g\). For each combination of treatment, gene, and gene alteration, we model the mean treatment effect as
\begin{equation}
    \E{\ell}{
        \Delta_{\ell,j}
        \mid x_{\ell,g}^{\mathsmaller{(m)}} = 0}
        \defeq \gamma_j, \qquad
    \E{\ell}{
        \Delta_{\ell,j}
        \mid x_{\ell,g}^{\mathsmaller{(m)}} = 1}
        \defeq
        \gamma_j + \delta_{j,g,m}.
\label{eq:gene-effect}
\end{equation}
Here, \(\gamma_j\) is the average treatment effect among PDX lines without alteration type \(m\) in gene \(g\), while \(\delta_{j,g,m}\) encodes the association between treatment response and the external genomic covariate \(x_{\ell,g}^{\mathsmaller{(m)}}\). We estimate \(\delta_{j,g,m}\) using the linear regression
\begin{equation}
    \Delta_{\ell,j}
    =
    \gamma_j
    +
    \delta_{j,g,m} x_{\ell,g}^{\mathsmaller{(m)}}
    +
    \epsilon_{\ell}.
    \label{eq:treatment-gene-association-regression}
\end{equation}
The regression is fit by ordinary least squares over PDX lines \(\ell\) for which treatment \(j\), untreated control outcomes, and alteration status for gene \(g\) are available.

We also compute treatment effects for both the minimum tumor size and the mean tumor size over the mouse's observed tumor trajectory. In these analyses, \(\bar{s}_{\ell,j}\) and \(\bar{s}_{\ell,0}\) denote the corresponding average tumor size response among treated and untreated mice from PDX line \(\ell\). For lifespan, larger values of \(\Delta_{\ell,j}\) indicate longer survival under treatment. For minimum and mean tumor size, smaller values indicate lower tumor burden under treatment.

\subsubsection{Agreement across genomic associations}
We compare observed and generated association coefficients across the full collection of fitted regressions. For a given treatment \(j\), gene \(g\), alteration type \(m\), and response \(r\), let \(\widehat{\delta}_{j,g,m,r}^{\mathsmaller{\mathrm{obs}}}\) denote the coefficient estimated from the observed dataset, and let \(\widehat{\delta}_{j,g,m,r}^{\mathsmaller{\mathrm{gen}}}\) denote the corresponding coefficient estimated from the generated dataset. This comparison assesses whether the generated outcomes preserve all the gene association patterns estimated from the observed PDX screen.

Because both coordinates are estimated association coefficients, we fit the calibration line using weighted orthogonal distance regression, an errors-in-variables method that accounts for uncertainty in both covariates and responses \citep{boggs1989orthogonal}. For each fixed alteration type \(m\) and response \(r\), let \(\mathcal{K}_{m,r}\) denote the collection of treatment-gene pairs included in the corresponding panel. We use these coefficient pairs to fit
\begin{equation}
    \widehat{\delta}_{j,g,m,r}^{\mathsmaller{\mathrm{gen}}}
    =
    \rho_{m,r}
    \widehat{\delta}_{j,g,m,r}^{\mathsmaller{\mathrm{obs}}}
    +
    \varepsilon_{j,g,m,r},
    \qquad
    (j,g)\in\mathcal{K}_{m,r}.
\label{eq:coefficient-calibration-model}
\end{equation}
Here \(\rho_{m,r}\) measures the multiplicative calibration of generated association coefficients relative to observed association coefficients for alteration type \(m\) and response \(r\). The ideal value is \(\rho_{m,r}=1\) as that would mean on average that the model preserves the associations in the original data. One interpretation of \(\rho_{m,r}\) is as an extension to a Pearson correlation coefficient. Pearson correlation is equivalent to an ordinary least-squares slope after centering and standardizing both variables, whereas \(\rho_{m,r}\) is an uncentered weighted orthogonal-regression slope that incorporates uncertainty in both coefficient estimates and need not lie in \([-1,1]\). 

\begin{figure}[ht]
    \centering
    \includegraphics[width=\linewidth, trim=0 0.25cm 0 0.25cm,clip]{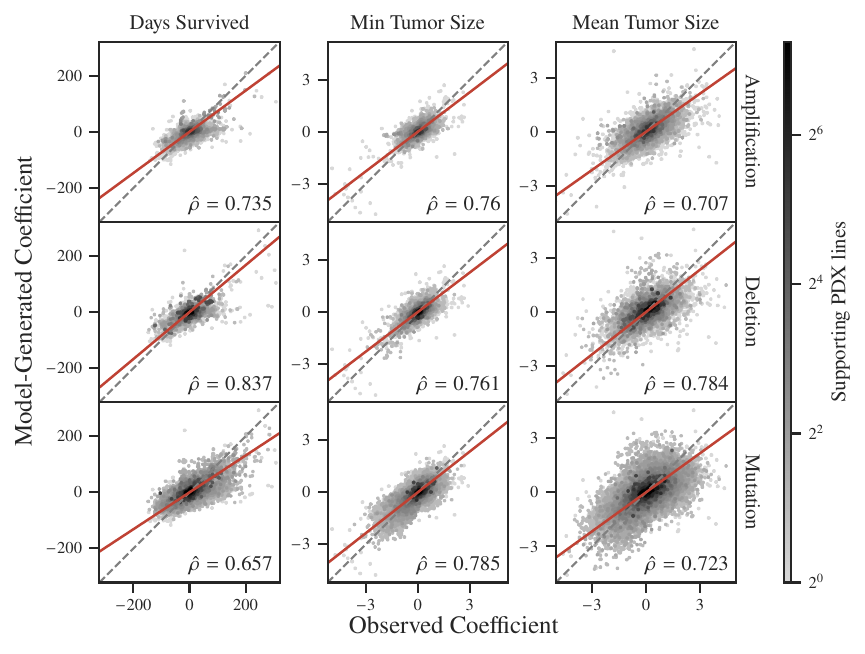}
    \caption{Each panel plot \(\widehat{\delta}_{j,g,m,r}^{\mathsmaller{\mathrm{gen}}}\) against \(\widehat{\delta}_{j,g,m,r}^{\mathsmaller{\mathrm{obs}}}\) for one alteration type and response. The positive $\hat{\rho}$ shows that the generated data preserve association directions, while slopes around $0.75$ indicate shrinkage of generated effects toward zero. In all plots the dashed line is the ideal $y=x$ line and the red line shows the regression fit.}
    \label{fig:allDrugsAllCoef}
\end{figure}

\Cref{fig:allDrugsAllCoef} plots \(\widehat{\delta}_{j,g,m,r}^{\mathsmaller{\mathrm{gen}}}\) against \(\widehat{\delta}_{j,g,m,r}^{\mathsmaller{\mathrm{obs}}}\) separately for each alteration type and response. All nine fitted calibration slopes \(\widehat{\rho}_{m,r}\) are positive, ranging from \(0.657\) to \(0.837\), with a mean of \(0.750\). Because every slope is below one, the generated coefficients tend to be closer to zero than the observed coefficients, with a mean fitted magnitude equal to \(75.0\%\) of the observed magnitude. Nevertheless, the \(95\%\) confidence intervals for the observed and generated coefficients overlap for \(99.6\%\) of all gene associations in the dataset, and the generated point estimate lies within the observed \(95\%\) confidence interval in \(97.4\%\) of cases. Thus, the generated data generally reproduce the association patterns estimated from the observed data within their reported uncertainty, while yielding point estimates of smaller magnitude.

\subsection{Repositioning candidates for untested PDX-treatment combinations}
A key motivation for \modelname is its ability to predict outcomes for PDX-treatment combinations not observed in the original screen. We use this capability to prioritize candidate future experiments. Eligibility is restricted to PDX lines empirically classified as having progressive disease under no treatment. For each eligible line, an untested treatment is considered a candidate if complete response has the highest posterior predictive probability among the mRECIST categories. Candidates are ranked by their predicted additional lifespan, measured in days relative to the untreated outcome for the same PDX line.

\Cref{tab:tflfpdx-unobserved-pair-ranking} presents the ten highest ranked candidates. Predicted death time under each condition is the median of its posterior predictive distribution. The accompanying 90\% posterior predictive interval (PPI) is computed from the posterior predictive distribution of the difference in death times. The untreated mRECIST classification is computed from the observed data, whereas the treated classification is inferred from posterior predictive tumor trajectories. For PDX lines empirically classified as having progressive disease when untreated, \(\Pr(\mathrm{PD}\to\mathrm{CR})\) denotes the posterior predictive probability of complete response under the candidate treatment. All ten candidates have positive predicted additional lifespan, although the 90\% PPIs include zero for four candidates. We note that the treatments LEE011, binimetinib, and encorafenib each appear more than once.

\begin{table} 
\caption{The ten highest ranked untested PDX-treatment combinations identified as candidate future experiments. Each candidate pairs a PDX line empirically classified as having progressive disease under no treatment with a treatment for which complete response receives the greatest posterior predictive probability. Candidates are ordered by predicted additional lifespan.}
\label{tab:tflfpdx-unobserved-pair-ranking} 
\begin{tabular}{rllll}
\toprule
Rank & PDX line & Treatment & Additional lifespan [90\% PPI] & $\Pr(\mathrm{PD}\to\mathrm{CR})$ \\
\midrule
1 & X-3483 / CM & LEE011 + binimetinib & 215 [15, 388] & 1.00 \\
2 & X-3746 / CM & LEE011 + binimetinib & 176 [4, 352] & 1.00 \\
3 & X-4644 / CM & LEE011 + binimetinib & 166 [3, 355] & 0.76 \\
4 & X-1228 / NSCLC & encorafenib + binimetinib & 92 [-3, 232] & 0.99 \\
5 & X-3483 / CM & BKM120 + binimetinib & 88 [5, 204] & 0.51 \\
6 & X-1228 / NSCLC & LEE011 + encorafenib & 87 [-3, 198] & 0.53 \\
7 & X-4567 / BRCA & BKM120 + encorafenib & 81 [-3, 203] & 0.75 \\
8 & X-3483 / CM & gemcitabine-50mpk & 79 [7, 152] & 0.95 \\
9 & X-4567 / BRCA & LEE011 + encorafenib & 78 [-2, 189] & 0.75 \\
10 & X-4567 / BRCA & LEE011 + binimetinib & 78 [1, 179] & 0.50 \\
\bottomrule
\end{tabular}
\end{table}

\modelname predicts that the top three candidates, X-3483/CM, X-3746/CM, and X-4644/CM, will respond to a combinations therapy of LEE011 and binimetinib. In our data, all three PDX lines carry a BRAF V600E mutation. Although binimetinib does not directly inhibit BRAF, it inhibits MEK1/2 downstream of BRAF in the MAPK pathway, making MEK inhibition biologically plausible in BRAF-mutant tumors~\citep{solit2006braf,Tran2020-ng}. These same three candidates also show alterations in the CDK4/6 regulatory axis. X-3483/CM has a partial deletion of CDKN2B, and X-3746/CM has a mutation in CDKN2A. Both genes encode INK4-family tumor suppressors that regulate CDK4/6 activity~\citep{ruas1998p16}. X-4644/CM has an amplification of CDK6, which LEE011 directly targets~\citep{Tripathy2017-mn}. Thus, the highest-ranked candidates are consistent with a biological hypothesis involving concurrent MAPK-pathway activation and CDK4/6-axis dysregulation. This shows that the model's predictions provide biologically coherent hypotheses for future experimentation.

\section{Conclusion}
\label{sec:conclusion}
This paper introduced \modelname, a hierarchical Bayesian model for analyzing longitudinal PDX data that jointly models tumor progression and survival under treatment. \modelname combines spline-based tumor trajectories, low-rank sharing across PDX lines and treatments, and a survival model in which latent tumor history contributes to death risk through a learned convolution kernel. This construction is designed for sparse PDX screens, where many PDX-treatment combinations are unobserved and where tumor measurements are informatively truncated by death. Across held-out and partially observed prediction tasks, \modelname improves survival prediction relative to standard survival baselines and produces competitive tumor response predictions that can update as early tumor measurements are incorporated.

\modelname links treatment response to survival through the latent tumor trajectory. The learned convolution kernel identifies the lag window over which previous tumor burden is most associated with current death risk, while the latent representations capture heterogeneity in response across tumors and treatments. Although genomic covariates are not used in fitting the model, posterior predictive outcomes preserve biologically meaningful treatment-gene response patterns and yield coherent hypotheses for untested PDX-treatment combinations. Thus, the model's latent structure supports both prediction and biological interrogation.

Several limitations point to natural extensions. \modelname currently learns response structure from tumor and survival outcomes alone, without incorporating external covariates such as mutation status, copy-number alterations, cancer type, treatment targets, or known pathway relationships. Adding these sources of information could improve prediction for PDX lines and treatments with few observations and would be essential for settings involving new PDX lines or new therapies. The current framework also ranks candidate experiments using posterior predictive summaries, but a full experimental design strategy would require decision rules that account for posterior uncertainty, validation cost, and the relative consequences of false positive and false negative recommendations. Future work should therefore integrate biological covariates into the latent representation, develop uncertainty aware criteria for experimental prioritization and early stopping, and validate the resulting candidates in independent PDX screens or prospective experiments.

Taken together, \modelname provides a unified probabilistic framework for sparse PDX screens, linking longitudinal tumor response, survival, and treatment heterogeneity within a single model. Using latent representations of tumor response and survival, the model provides a foundation for forecasting treatment response, prioritizing follow-up experiments, and developing future tools for PDX study design.

\section{Acknowledgments}
\label{sec:ack}
We are grateful to Olivia Tsang, Ankur Garg, Amber Lee, and Christopher Wolfram for their thoughtful comments on figures and prose.
WT is supported by the NIH/NCI (R37 CA271186, U54 CA274492, P30 CA008748), Break Through Cancer, the Fund for Innovation in Cancer Informatics, the Cancer AI Alliance, the Tow Center for Developmental Oncology, and the Maurice Campbell Initiative at Memorial Sloan Kettering Cancer Center.

\ifAOAS
    \bibliographystyle{preamble_aoas/imsart-nameyear}
\else
    \bibliographystyle{plainnat}
\fi
\bibliography{references}

@article{Cox1972,
 ISSN = {00359246},
 URL = {http://www.jstor.org/stable/2985181},
 author = {D. R. Cox},
 journal = {Journal of the Royal Statistical Society. Series B (Methodological)},
 number = {2},
 pages = {187--220},
 publisher = {[Royal Statistical Society, Oxford University Press]},
 title = {Regression Models and Life-Tables},
 urldate = {2026-03-18},
 volume = {34},
 year = {1972}
}

@ARTICLE{Tripathy2017-mn,
  title     = "Ribociclib ({LEE011)}: Mechanism of action and clinical impact
               of this selective cyclin-dependent kinase 4/6 inhibitor in
               various solid tumors",
  author    = "Tripathy, Debu and Bardia, Aditya and Sellers, William R",
  journal   = "Clin. Cancer Res.",
  publisher = "American Association for Cancer Research (AACR)",
  volume    =  23,
  number    =  13,
  pages     = "3251--3262",
  year      =  2017,
  language  = "en"
}

@Article{Gelman2014,
author={Gelman, Andrew
and Hwang, Jessica
and Vehtari, Aki},
title={Understanding predictive information criteria for Bayesian models},
journal={Statistics and Computing},
year={2014},
month={Nov},
day={01},
volume={24},
number={6},
pages={997-1016},
issn={1573-1375},
doi={10.1007/s11222-013-9416-2},
url={https://doi.org/10.1007/s11222-013-9416-2}
}

@Article{Gao2015,
author={Gao, Hui
and Korn, Joshua M.
and Ferretti, St{\'e}phane
and Monahan, John E.
and Wang, Youzhen
and Singh, Mallika
and Zhang, Chao
and Schnell, Christian
and Yang, Guizhi
and Zhang, Yun
and Balbin, O. Alejandro
and Barbe, St{\'e}phanie
and Cai, Hongbo
and Casey, Fergal
and Chatterjee, Susmita
and Chiang, Derek Y.
and Chuai, Shannon
and Cogan, Shawn M.
and Collins, Scott D.
and Dammassa, Ernesta
and Ebel, Nicolas
and Embry, Millicent
and Green, John
and Kauffmann, Audrey
and Kowal, Colleen
and Leary, Rebecca J.
and Lehar, Joseph
and Liang, Ying
and Loo, Alice
and Lorenzana, Edward
and Robert McDonald, E.
and McLaughlin, Margaret E.
and Merkin, Jason
and Meyer, Ronald
and Naylor, Tara L.
and Patawaran, Montesa
and Reddy, Anupama
and R{\"o}elli, Claudia
and Ruddy, David A.
and Salangsang, Fernando
and Santacroce, Francesca
and Singh, Angad P.
and Tang, Yan
and Tinetto, Walter
and Tobler, Sonja
and Velazquez, Roberto
and Venkatesan, Kavitha
and Von Arx, Fabian
and Wang, Hui Qin
and Wang, Zongyao
and Wiesmann, Marion
and Wyss, Daniel
and Xu, Fiona
and Bitter, Hans
and Atadja, Peter
and Lees, Emma
and Hofmann, Francesco
and Li, En
and Keen, Nicholas
and Cozens, Robert
and Jensen, Michael Rugaard
and Pryer, Nancy K.
and Williams, Juliet A.
and Sellers, William R.},
title={High-throughput screening using patient-derived tumor xenografts to predict clinical trial drug response},
journal={Nature Medicine},
year={2015},
month={Nov},
day={01},
volume={21},
number={11},
pages={1318-1325},
issn={1546-170X},
doi={10.1038/nm.3954},
url={https://doi.org/10.1038/nm.3954}
}

@misc{Naguib2026,
author = {Naguib, Karim and Berché, Roger and Li, Lu and Bevan, Antonia and Khosla, Sajan and Davies, Jessica and Metcalfe, Paul},
year = {2026},
month = {07},
pages = {},
title = {PIONEER: Bayesian Joint Modelling of Mechanistic Tumour Growth and Time-to-Event Endpoints for Dynamic Prediction of Ongoing Oncology Trials},
doi = {10.48550/arXiv.2607.17908}
}

@article{Pan2014,
author = {Pan, Jianxin and Bao, Yanchun and Dai, Hongsheng and Fang, Hong-Bin},
title = {Joint longitudinal and survival-cure models in tumour xenograft experiments},
journal = {Statistics in Medicine},
volume = {33},
number = {18},
pages = {3229-3240},
doi = {https://doi.org/10.1002/sim.6175},
url = {https://onlinelibrary.wiley.com/doi/abs/10.1002/sim.6175},
eprint = {https://onlinelibrary.wiley.com/doi/pdf/10.1002/sim.6175},
year = {2014}
}

@article{Sylvestre2009,
author = {Sylvestre, Marie-Pierre and Abrahamowicz, Michal},
title = {Flexible modeling of the cumulative effects of time-dependent exposures on the hazard},
journal = {Statistics in Medicine},
volume = {28},
number = {27},
pages = {3437-3453},
doi = {https://doi.org/10.1002/sim.3701},
url = {https://onlinelibrary.wiley.com/doi/abs/10.1002/sim.3701},
eprint = {https://onlinelibrary.wiley.com/doi/pdf/10.1002/sim.3701},
year = {2009}
}

@article{aalen1989linear,
author = {Aalen, Odd O.},
title = {A linear regression model for the analysis of life times},
journal = {Statistics in Medicine},
volume = {8},
number = {8},
pages = {907-925},
doi = {https://doi.org/10.1002/sim.4780080803},
url = {https://onlinelibrary.wiley.com/doi/abs/10.1002/sim.4780080803},
eprint = {https://onlinelibrary.wiley.com/doi/pdf/10.1002/sim.4780080803},
year = {1989}
}

@incollection{aalen1980nonparametric,
author={Aalen, Odd},
editor={Klonecki, Witold
and Kozek, Andrzej
and Rosi{\'{n}}ski, Jan},
title={A Model for Nonparametric Regression Analysis of Counting Processes},
booktitle={Mathematical Statistics and Probability Theory},
year={1980},
publisher={Springer New York},
address={New York, NY},
pages={1-25},
isbn={978-1-4615-7397-5}
}

@article{tansey:etal:2022:bayesian-dose-response,
  author  = {Tansey, Wesley and Tosh, Christopher and Blei, David M.},
  title   = {A Bayesian Model of Dose--Response for Cancer Drug Studies},
  journal = {The Annals of Applied Statistics},
  year    = {2022},
  volume  = {16},
  number  = {2},
  pages   = {680--705},
}

@article{boggs1989orthogonal,
  title={Orthogonal distance regression},
  author={Boggs, Paul T and  Rogers, Janet E},
  year={1989},
  publisher={US Department of Commerce, National Institute of Standards and Technology}
}

@article{Carroll1970,
author={Carroll, J. Douglas
and Chang, Jih-Jie},
title={Analysis of individual differences in multidimensional scaling via an n-way generalization of ``Eckart-Young'' decomposition},
journal={Psychometrika},
year={1970},
month={Sep},
day={01},
volume={35},
number={3},
pages={283-319},
issn={1860-0980},
doi={10.1007/BF02310791},
url={https://doi.org/10.1007/BF02310791}
}

@article{kolda2009,
author = {Kolda, Tamara G. and Bader, Brett W.},
title = {Tensor Decompositions and Applications},
journal = {SIAM Review},
volume = {51},
number = {3},
pages = {455-500},
year = {2009},
doi = {10.1137/07070111X},

URL = {     
        https://doi.org/10.1137/07070111X
},
eprint = { 
    
        https://doi.org/10.1137/07070111X
}
}

@article{ChenG16,
  author       = {Tianqi Chen and
                  Carlos Guestrin},
  title        = {XGBoost: {A} Scalable Tree Boosting System},
  journal      = {CoRR},
  volume       = {abs/1603.02754},
  year         = {2016},
  url          = {http://arxiv.org/abs/1603.02754},
  eprinttype   = {arXiv},
  eprint       = {1603.02754},
  bibsource    = {dblp computer science bibliography, https://dblp.org}
}

@ARTICLE{Royston2002-we,
  title     = "Flexible parametric proportional-hazards and proportional-odds
               models for censored survival data, with application to
               prognostic modelling and estimation of treatment effects",
  author    = "Royston, Patrick and Parmar, Mahesh K B",
  journal   = "Stat. Med.",
  publisher = "Wiley",
  volume    =  21,
  number    =  15,
  pages     = "2175--2197",
  year      =  2002,
  copyright = "http://onlinelibrary.wiley.com/termsAndConditions\#vor",
  language  = "en"
}

@article{Breslow72,
 ISSN = {00359246},
 URL = {http://www.jstor.org/stable/2985181},
 author = {Norman Breslow},
 journal = {Journal of the Royal Statistical Society. Series B (Methodological)},
 number = {2},
 pages = {216--217},
 publisher = {[Royal Statistical Society, Oxford University Press]},
 title = {Discussion of the paper by {D. R. Cox}},
 urldate = {2025-05-27},
 volume = {34},
 year = {1972}
}

@article{Claret2013,
author = {Claret, Laurent and Gupta, Manish and Han, Kelong and Joshi, Amita and Sarapa, Nenad and He, Jing and Powell, Bob and Bruno, René },
title = {Evaluation of Tumor-Size Response Metrics to Predict Overall Survival in Western and Chinese Patients With First-Line Metastatic Colorectal Cancer},
journal = {Journal of Clinical Oncology},
volume = {31},
number = {17},
pages = {2110-2114},
year = {2013},
doi = {10.1200/JCO.2012.45.0973},
note ={PMID: 23650411},
URL = {https://ascopubs.org/doi/abs/10.1200/JCO.2012.45.0973},
eprint = {https://ascopubs.org/doi/pdf/10.1200/JCO.2012.45.0973}
}

@article{Claret2009,
author = {Claret, Laurent and Girard, Pascal and Hoff, Paulo M. and Van Cutsem, Eric and Zuideveld, Klaas P. and Jorga, Karin and Fagerberg, Jan and Bruno, René },
title = {Model-Based Prediction of Phase III Overall Survival in Colorectal Cancer on the Basis of Phase II Tumor Dynamics},
journal = {Journal of Clinical Oncology},
volume = {27},
number = {25},
pages = {4103-4108},
year = {2009},
doi = {10.1200/JCO.2008.21.0807},
note ={PMID: 19636014},
URL = {https://ascopubs.org/doi/abs/10.1200/JCO.2008.21.0807},
eprint = {https://ascopubs.org/doi/pdf/10.1200/JCO.2008.21.0807}
}

@article{Wang2025,
   author = "Wang, Jane-Ling and Zhong, Qixian",
   title = "Joint Modeling of Longitudinal and Survival Data", 
   journal= "Annual Review of Statistics and Its Application",
   year = "2025",
   volume = "12",
   number = "Volume 12, 2025",
   pages = "449-476",
   doi = "https://doi.org/10.1146/annurev-statistics-112723-034334",
   url = "https://www.annualreviews.org/content/journals/10.1146/annurev-statistics-112723-034334",
   publisher = "Annual Reviews",
   issn = "2326-831X",
   type = "Journal Article",
  }

@book{deboor1978,
  address = {New York},
  author = {De Boor, Carl},
  publisher = {Springer Verlag},
  title = {A Practical Guide to Splines},
  year = 1978
}

@article{Gupta2025,
    author = {Gupta, Shruti and Mohani, Vikash K and Ghislat, Ghita and Ballester, Pedro J and Ahmad, Shandar},
    title = {A pan-cancer, pan-treatment model for predicting drug responses from patient-derived xenografts},
    journal = {NAR Genomics and Bioinformatics},
    volume = {7},
    number = {3},
    pages = {lqaf111},
    year = {2025},
    month = {09},
    issn = {2631-9268},
    doi = {10.1093/nargab/lqaf111},
    url = {https://doi.org/10.1093/nargab/lqaf111},
    eprint = {https://academic.oup.com/nargab/article-pdf/7/3/lqaf111/64147245/lqaf111.pdf},
}

@Article{Bertsimas2023,
author={Bertsimas, Dimitris
and Pawlowski, Colin},
title={Tensor completion with noisy side information},
journal={Machine Learning},
year={2023},
month={Oct},
day={01},
volume={112},
number={10},
pages={3945-3976},
issn={1573-0565},
doi={10.1007/s10994-023-06338-5},
url={https://doi.org/10.1007/s10994-023-06338-5}
}

@ARTICLE{Ren2022,
  title    = "De novo prediction of cell-drug sensitivities using deep
              learning-based graph regularized matrix factorization",
  author   = "Ren, Shuangxia and Tao, Yifeng and Yu, Ke and Xue, Yifan and
              Schwartz, Russell and Lu, Xinghua",
  journal  = "Pac. Symp. Biocomput.",
  volume   =  27,
  pages    = "278--289",
  year     =  2022,
  language = "en"
}

@article{Mourragui2021,
author = {Soufiane M. C. Mourragui  and Marco Loog  and Daniel J. Vis  and Kat Moore  and Anna G. Manjon  and Mark A. van de Wiel  and Marcel J. T. Reinders  and Lodewyk F. A. Wessels },
title = {Predicting patient response with models trained on cell lines and patient-derived xenografts by nonlinear transfer learning},
journal = {Proceedings of the National Academy of Sciences},
volume = {118},
number = {49},
pages = {e2106682118},
year = {2021},
doi = {10.1073/pnas.2106682118},
URL = {https://www.pnas.org/doi/abs/10.1073/pnas.2106682118},
eprint = {https://www.pnas.org/doi/pdf/10.1073/pnas.2106682118}}

@Article{Armingol2022,
author={Armingol, Erick
and Baghdassarian, Hratch M.
and Martino, Cameron
and Perez-Lopez, Araceli
and Aamodt, Caitlin
and Knight, Rob
and Lewis, Nathan E.},
title={Context-aware deconvolution of cell--cell communication with Tensor-cell2cell},
journal={Nature Communications},
year={2022},
month={Jun},
day={27},
volume={13},
number={1},
pages={3665},
issn={2041-1723},
doi={10.1038/s41467-022-31369-2},
url={https://doi.org/10.1038/s41467-022-31369-2}
}

@inproceedings{Schein2015,
author = {Schein, Aaron and Paisley, John and Blei, David M. and Wallach, Hanna},
title = {Bayesian Poisson Tensor Factorization for Inferring Multilateral Relations from Sparse Dyadic Event Counts},
year = {2015},
isbn = {9781450336642},
publisher = {Association for Computing Machinery},
address = {New York, NY, USA},
url = {https://doi.org/10.1145/2783258.2783414},
doi = {10.1145/2783258.2783414},
booktitle = {Proceedings of the 21th ACM SIGKDD International Conference on Knowledge Discovery and Data Mining},
pages = {1045–1054},
numpages = {10},
location = {Sydney, NSW, Australia},
series = {KDD '15}
}

@ARTICLE{Zhou2023,
    
AUTHOR={Zhou, Jie  and Jiang, Xun  and Xia, H. Amy  and Hobbs, Brian P.  and Wei, Peng },
           
TITLE={Landmark mediation survival analysis using longitudinal surrogate},
          
JOURNAL={Frontiers in Oncology},
          
VOLUME={Volume 12 - 2022},
  
YEAR={2023},
  
URL={https://www.frontiersin.org/journals/oncology/articles/10.3389/fonc.2022.999324},
  
DOI={10.3389/fonc.2022.999324},
  
ISSN={2234-943X}}

@article{Harshman1970,
  author = {Harshman, R. A.},
  description = {papers on three-mode analysis},
  journal = {UCLA Working Papers in Phonetics},
  pages = {1--84},
  title = {{F}oundations of the {P}{A}{R}{A}{F}{A}{C} procedure: {M}odels and conditions for an "explanatory" multi-modal factor analysis},
  volume = 16,
  year = 1970
}

@article{jarada_review_2020,
author={Jarada, Tamer N.
and Rokne, Jon G.
and Alhajj, Reda},
title={A review of computational drug repositioning: strategies, approaches, opportunities, challenges, and directions},
journal={Journal of Cheminformatics},
year={2020},
month={Jul},
day={22},
volume={12},
number={1},
pages={46},
issn={1758-2946},
doi={10.1186/s13321-020-00450-7},
url={https://doi.org/10.1186/s13321-020-00450-7}
}

@article{peska_drug-target_2017,
title = {Drug-target interaction prediction: A Bayesian ranking approach},
journal = {Computer Methods and Programs in Biomedicine},
volume = {152},
pages = {15-21},
year = {2017},
issn = {0169-2607},
doi = {https://doi.org/10.1016/j.cmpb.2017.09.003},
url = {https://www.sciencedirect.com/science/article/pii/S0169260717303413},
author = {Ladislav Peska and Krisztian Buza and Júlia Koller}
}

@ARTICLE{Gasparrini2010,
  title     = "Distributed lag non-linear models",
  author    = "Gasparrini, A and Armstrong, B and Kenward, M G",
  journal   = "Stat. Med.",
  publisher = "Wiley",
  volume    =  29,
  number    =  21,
  pages     = "2224--2234",
  year      =  2010,
  copyright = "http://onlinelibrary.wiley.com/termsAndConditions\#vor",
  language  = "en"
}

@article{Wang2021,
    author = {Wang, Tianduanyi and Szedmak, Sandor and Wang, Haishan and Aittokallio, Tero and Pahikkala, Tapio and Cichonska, Anna and Rousu, Juho},
    title = {Modeling drug combination effects via latent tensor reconstruction},
    journal = {Bioinformatics},
    volume = {37},
    number = {Supplement_1},
    pages = {i93-i101},
    year = {2021},
    month = {07},
    issn = {1367-4803},
    doi = {10.1093/bioinformatics/btab308},
    url = {https://doi.org/10.1093/bioinformatics/btab308},
    eprint = {https://academic.oup.com/bioinformatics/article-pdf/37/Supplement_1/i93/50694237/btab308.pdf},
}

@article{Suphavilai2018,
    author = {Suphavilai, Chayaporn and Bertrand, Denis and Nagarajan, Niranjan},
    title = {Predicting Cancer Drug Response using a Recommender System},
    journal = {Bioinformatics},
    volume = {34},
    number = {22},
    pages = {3907-3914},
    year = {2018},
    month = {11},
    issn = {1367-4803},
    doi = {10.1093/bioinformatics/bty452},
    url = {https://doi.org/10.1093/bioinformatics/bty452},
    eprint = {https://academic.oup.com/bioinformatics/article-pdf/34/22/3907/48920790/bioinformatics_34_22_3907.pdf},
}

@article{Ammaduddin2014,
    author = {Ammad-ud-din, Muhammad and Georgii, Elisabeth and G\"{o}nen, Mehmet and Laitinen, Tuomo and Kallioniemi, Olli and Wennerberg, Krister and Poso, Antti and Kaski, Samuel},
    title = {Integrative
and Personalized QSAR Analysis in Cancer
by Kernelized Bayesian Matrix Factorization},
    journal = {Journal of Chemical Information and Modeling},
    volume = {54},
    number = {8},
    pages = {2347-2359},
    year = {2014},
    month = {07},
    issn = {1549-9596},
    doi = {10.1021/ci500152b},
    url = {https://doi.org/10.1021/ci500152b},
    eprint = {https://pubs.acs.org/jcisd8/article-pdf/54/8/2347/10064548/ci500152b.pdf},
}

@Article{Kurilov2020,
author={Kurilov, Roman
and Haibe-Kains, Benjamin
and Brors, Benedikt},
title={Assessment of modelling strategies for drug response prediction in cell lines and xenografts},
journal={Scientific Reports},
year={2020},
month={Feb},
day={18},
volume={10},
number={1},
pages={2849},
issn={2045-2322},
doi={10.1038/s41598-020-59656-2},
url={https://doi.org/10.1038/s41598-020-59656-2}
}

@ARTICLE{OConnell2021,
  title     = "Metabolic biomarkers for the early detection of cancer cachexia",
  author    = "O'Connell, Thomas M and Golzarri-Arroyo, Lilian and Pin,
               Fabrizio and Barreto, Rafael and Dickinson, Stephanie L and
               Couch, Marion E and Bonetto, Andrea",
  journal   = "Front. Cell Dev. Biol.",
  publisher = "Frontiers Media SA",
  volume    =  9,
  pages     = "720096",
  year      =  2021,
  copyright = "https://creativecommons.org/licenses/by/4.0/",
  language  = "en"
}

@ARTICLE{Tardivon2019,
  title     = "Association between tumor size kinetics and survival in patients
               with urothelial carcinoma treated with atezolizumab: Implication
               for patient follow-up",
  author    = "Tardivon, Coralie and Desm{\'e}e, Sol{\`e}ne and Kerioui, Marion
               and Bruno, Ren{\'e} and Wu, Benjamin and Mentr{\'e}, France and
               Mercier, Fran{\c c}ois and Guedj, J{\'e}r{\'e}mie",
  journal   = "Clin. Pharmacol. Ther.",
  publisher = "Wiley",
  volume    =  106,
  number    =  4,
  pages     = "810--820",
  year      =  2019,
  copyright = "http://onlinelibrary.wiley.com/termsAndConditions\#vor",
  language  = "en"
}

@ARTICLE{Mauff2017,
  title     = "Extension of the association structure in joint models to
               include weighted cumulative effects",
  author    = "Mauff, Katya and Steyerberg, Ewout W and Nijpels, Giel and van
               der Heijden, Amber A W A and Rizopoulos, Dimitris",
  journal   = "Stat. Med.",
  publisher = "Wiley",
  volume    =  36,
  number    =  23,
  pages     = "3746--3759",
  year      =  2017,
  copyright = "http://onlinelibrary.wiley.com/termsAndConditions\#vor",
  language  = "en"
}

@article{zhang_drimc_2020,
    author = {Zhang, Wenjuan and Xu, Hunan and Li, Xiaozhong and Gao, Qiang and Wang, Lin},
    title = {DRIMC: an improved drug repositioning approach using Bayesian inductive matrix completion},
    journal = {Bioinformatics},
    volume = {36},
    number = {9},
    pages = {2839-2847},
    year = {2020},
    month = {05},
    issn = {1367-4803},
    doi = {10.1093/bioinformatics/btaa062},
    url = {https://doi.org/10.1093/bioinformatics/btaa062},
    eprint = {https://academic.oup.com/bioinformatics/article-pdf/36/9/2839/48986273/bioinformatics_36_9_2839.pdf},
}

@article{yang_drugdisease_2014,
    author = {Yang, Jihong and Li, Zheng and Fan, Xiaohui and Cheng, Yiyu},
    title = {Drug–Disease
Association and Drug-Repositioning
Predictions in Complex Diseases Using Causal Inference–Probabilistic
Matrix Factorization},
    journal = {Journal of Chemical Information and Modeling},
    volume = {54},
    number = {9},
    pages = {2562-2569},
    year = {2014},
    month = {08},
    issn = {1549-9596},
    doi = {10.1021/ci500340n},
    url = {https://doi.org/10.1021/ci500340n},
    eprint = {https://pubs.acs.org/jcisd8/article-pdf/54/9/2562/10078179/ci500340n.pdf},
}

@article{rizopoulos_dynamic_2017,
author = {Rizopoulos, Dimitris and Molenberghs, Geert and Lesaffre, Emmanuel M.E.H.},
title = {Dynamic predictions with time-dependent covariates in survival analysis using joint modeling and landmarking},
journal = {Biometrical Journal},
volume = {59},
number = {6},
pages = {1261-1276},
doi = {https://doi.org/10.1002/bimj.201600238},
url = {https://onlinelibrary.wiley.com/doi/abs/10.1002/bimj.201600238},
eprint = {https://onlinelibrary.wiley.com/doi/pdf/10.1002/bimj.201600238},
year = {2017}
}

@article{li_dynamic_2019,
author = {Li, Kan and Luo, Sheng},
title = {Functional joint model for longitudinal and time-to-event data: an application to Alzheimer's disease},
journal = {Statistics in Medicine},
volume = {36},
number = {22},
pages = {3560-3572},
doi = {https://doi.org/10.1002/sim.7381},
url = {https://onlinelibrary.wiley.com/doi/abs/10.1002/sim.7381},
eprint = {https://onlinelibrary.wiley.com/doi/pdf/10.1002/sim.7381},
year = {2017}
}

@article{alvares_tractable_2021,
author = {Alvares, Danilo and Rubio, Francisco J.},
title = {A tractable Bayesian joint model for longitudinal and survival data},
journal = {Statistics in Medicine},
volume = {40},
number = {19},
pages = {4213-4229},
doi = {https://doi.org/10.1002/sim.9024},
url = {https://onlinelibrary.wiley.com/doi/abs/10.1002/sim.9024},
eprint = {https://onlinelibrary.wiley.com/doi/pdf/10.1002/sim.9024},
year = {2021}
}

@article{ibrahim_basic_2010,
author = {Ibrahim, Joseph G. and Chu, Haitao and Chen, Liddy M. },
title = {Basic Concepts and Methods for Joint Models of Longitudinal and Survival Data},
journal = {Journal of Clinical Oncology},
volume = {28},
number = {16},
pages = {2796-2801},
year = {2010},
doi = {10.1200/JCO.2009.25.0654},

note ={PMID: 20439643},

URL = {https://ascopubs.org/doi/abs/10.1200/JCO.2009.25.0654},
eprint = {https://ascopubs.org/doi/pdf/10.1200/JCO.2009.25.0654}
}

@article{tentler:etal:2012:pdx-cancer-review,
  title={Patient-derived tumour xenografts as models for oncology drug development},
  author={Tentler, John J and Tan, Aik Choon and Weekes, Colin D and Jimeno, Antonio and Leong, Stephen and Pitts, Todd M and Arcaroli, John J and Messersmith, Wells A and Eckhardt, S Gail},
  journal={Nature Reviews Clinical oncology},
  volume={9},
  number={6},
  pages={338--350},
  year={2012},
  publisher={Nature Publishing Group UK London}
}

@article{matschos:etal:2021:hroc-pdx-biobank,
  title={The {HROC}-Xenobank—A high quality assured PDX biobank of> 100 individual colorectal cancer models},
  author={Matschos, Stephanie and B{\"u}rtin, Florian and Kdimati, Said and Radefeldt, Mandy and Krake, Susann and Prall, Friedrich and Engel, Nadja and Krohn, Mathias and Micheel, Bianca and Kreutzer, Michael and others},
  journal={Cancers},
  volume={13},
  number={23},
  pages={5882},
  year={2021},
  publisher={MDPI}
}

@article{akemi:etal:2025:pediatric-pdx-biobank,
  title={Establishing a pediatric solid tumor PDX biobank for precision oncology research},
  author={Akemi Kido, Larissa and Rodrigues Marusco, Milena and Aparecida da Silva, Ellen and Do Carmo, La{\'\i}s and Beatriz Teodoro Borges, Ana and Luz Torres Silva, Felipe and Silveira Ruas, Juliana and Giomo de Lima, Dieila and de Abreu Fernandes, Larissa and Maia Martin Daiggi, Camila and others},
  journal={Cancer Biology \& Therapy},
  volume={26},
  number={1},
  pages={2541974},
  year={2025},
  publisher={Taylor \& Francis}
}

@article{meric:etal:2024:pdxnet,
  title={Assessment of patient-derived xenograft growth and antitumor activity: the NCI PDXNet consensus recommendations},
  author={Meric-Bernstam, Funda and Lloyd, Michael W and Koc, Soner and Evrard, Yvonne A and McShane, Lisa M and Lewis, Michael T and Evans, Kurt W and Li, Dali and Rubinstein, Lawrence and Welm, Alana and others},
  journal={Molecular Cancer Therapeutics},
  volume={23},
  number={7},
  pages={924--938},
  year={2024},
  publisher={American Association for Cancer Research}
}

@article{haslam:etal:2021:eligibility-response-targeted-therapy,
  title={Updated estimates of eligibility for and response to genome-targeted oncology drugs among US cancer patients, 2006-2020},
  author={Haslam, A and Kim, MS and Prasad, V},
  journal={Annals of Oncology},
  volume={32},
  number={7},
  pages={926--932},
  year={2021},
  publisher={Elsevier}
}

@misc{Stan,
Author = {{Stan Development Team}},
Title = {Stan Reference Manual, version 2.39},
url ={https://mc-stan.org},
Year = {2026}
}

@article{Solit2006braf,
  title     = "{BRAF} mutation predicts sensitivity to {MEK} inhibition",
  author    = "Solit, David B and Garraway, Levi A and Pratilas, Christine A
               and Sawai, Ayana and Getz, Gad and Basso, Andrea and Ye, Qing
               and Lobo, Jose M and She, Yuhong and Osman, Iman and Golub, Todd
               R and Sebolt-Leopold, Judith and Sellers, William R and Rosen,
               Neal",
  journal   = "Nature",
  publisher = "Springer Science and Business Media LLC",
  volume    =  439,
  number    =  7074,
  pages     = "358--362",
  year      =  2006,
  language  = "en"
}

@article{ruas1998p16,
title = {The p16INK4a/CDKN2A tumor suppressor and its relatives},
journal = {Biochimica et Biophysica Acta (BBA) - Reviews on Cancer},
volume = {1378},
number = {2},
pages = {F115-F177},
year = {1998},
issn = {0304-419X},
doi = {https://doi.org/10.1016/S0304-419X(98)00017-1},
url = {https://www.sciencedirect.com/science/article/pii/S0304419X98000171},
author = {Margarida Ruas and Gordon Peters}
}

@ARTICLE{Tran2020-ng,
  title     = "The discovery and development of binimetinib for the treatment
               of melanoma",
  author    = "Tran, Brian and Cohen, Mark S",
  journal   = "Expert Opin. Drug Discov.",
  publisher = "Informa UK Limited",
  volume    =  15,
  number    =  7,
  pages     = "745--754",
  year      =  2020,
  language  = "en"
}

@ARTICLE{Wu2010-zh,
  title     = "Statistical inference for tumor growth inhibition {T/C} ratio",
  author    = "Wu, Jianrong",
  journal   = "J. Biopharm. Stat.",
  publisher = "Informa UK Limited",
  volume    =  20,
  number    =  5,
  pages     = "954--964",
  year      =  2010,
  language  = "en"
}

@article{Delord2017,
    author = {Delord, Jean-Pierre and Robert, Caroline and Nyakas, Marta and McArthur, Grant A. and Kudchakar, Ragini and Mahipal, Amit and Yamada, Yasuhide and Sullivan, Ryan and Arance, Ana and Kefford, Richard F. and Carlino, Matteo S. and Hidalgo, Manuel and Gomez-Roca, Carlos and Michel, Daniela and Seroutou, Abdelkader and Aslanis, Vassilios and Caponigro, Giordano and Stuart, Darrin D. and Moutouh-de Parseval, Laure and Demuth, Tim and Dummer, Reinhard},
    title = {Phase I Dose-Escalation and -Expansion Study of the BRAF Inhibitor Encorafenib (LGX818) in Metastatic BRAF-Mutant Melanoma},
    journal = {Clinical Cancer Research},
    volume = {23},
    number = {18},
    pages = {5339-5348},
    year = {2017},
    month = {09},
    issn = {1078-0432},
    doi = {10.1158/1078-0432.CCR-16-2923},
    url = {https://doi.org/10.1158/1078-0432.CCR-16-2923},
    eprint = {https://aacrjournals.org/clincancerres/article-pdf/23/18/5339/3418958/5339.pdf},
}

@ARTICLE{Young2014,
  title     = "Loss of {CDKN2A} expression is a frequent event in primary
               invasive melanoma and correlates with sensitivity to the
               {CDK4/6} inhibitor {PD0332991} in melanoma cell lines",
  author    = "Young, Richard J and Waldeck, Kelly and Martin, Claire and Foo,
               Jung H and Cameron, Donald P and Kirby, Laura and Do, Hongdo and
               Mitchell, Catherine and Cullinane, Carleen and Liu, Wendy and
               Fox, Stephen B and Dutton-Regester, Ken and Hayward, Nicholas K
               and Jene, Nicholas and Dobrovic, Alexander and Pearson, Richard
               B and Christensen, James G and Randolph, Sophia and McArthur,
               Grant A and Sheppard, Karen E",
  journal   = "Pigment Cell Melanoma Res.",
  publisher = "Wiley",
  volume    =  27,
  number    =  4,
  pages     = "590--600",
  month     =  jul,
  year      =  2014,
  copyright = "http://onlinelibrary.wiley.com/termsAndConditions\#vor",
  language  = "en"
}

@Article{Saridogan2023,
author={Saridogan, Turcin
and Akcakanat, Argun
and Zhao, Ming
and Evans, Kurt W.
and Yuca, Erkan
and Scott, Stephen
and Kirby, Bryce P.
and Zheng, Xiaofeng
and Ha, Min Jin
and Chen, Huiqin
and Ng, Patrick K. S.
and DiPeri, Timothy P.
and Mills, Gordon B.
and Rodon Ahnert, Jordi
and Damodaran, Senthil
and Meric-Bernstam, Funda},
title={Efficacy of futibatinib, an irreversible fibroblast growth factor receptor inhibitor, in FGFR-altered breast cancer},
journal={Scientific Reports},
year={2023},
month={Nov},
day={18},
volume={13},
number={1},
pages={20223},
issn={2045-2322},
doi={10.1038/s41598-023-46586-y},
url={https://doi.org/10.1038/s41598-023-46586-y}
}

@book{Brooks_2011,
   title={Handbook of Markov Chain Monte Carlo},
   ISBN={9780429138508},
   url={http://dx.doi.org/10.1201/b10905},
   DOI={10.1201/b10905},
   publisher={Chapman and Hall/CRC},
   author={Brooks, Steve and Gelman, Andrew and Jones, Galin and Meng, Xiao-Li},
   year={2011},
}

@article{hoffman14a,
  author  = {Matthew D. Hoffman and Andrew Gelman},
  title   = {The No-U-Turn Sampler: Adaptively Setting Path Lengths in Hamiltonian Monte Carlo},
  journal = {Journal of Machine Learning Research},
  year    = {2014},
  volume  = {15},
  number  = {47},
  pages   = {1593--1623},
  url     = {http://jmlr.org/papers/v15/hoffman14a.html}
}

\newpage
\appendix
\section{Additional Model and Evaluation Details}
\label{sec:appendix}

\subsection{Baseline and hyperparameter tuning information}
\label{ref:HyperparamTuning}
The implementations for multinomial regression and the multilayer perceptron come from the \texttt{Scikit-Learn} Python library. The implementation of XGBoost comes from the \texttt{XGBoost} Python library and the implementations for the Royston-Parmar model and Cox proportional hazards model come from the \texttt{Lifelines} Python library. Cross-validation was done on a held-out test set and details about hyperparameter grids chosen for tuning can be found in \Cref{tab:baseline-hyperparameter-tuning}. We select the mRECIST baseline hyperparameters to maximize validation accuracy on the tuning split. For the survival baselines the selection criterion was highest mean log probability.

\begin{table}[ht]
\centering
\scriptsize
\setlength{\tabcolsep}{3pt}
\renewcommand{\arraystretch}{1.15}
\caption{Hyperparameter tuning grids for the mRECIST and survival baselines.}
\label{tab:baseline-hyperparameter-tuning}
\begin{tabular}{
    @{}
    >{\raggedright\arraybackslash}p{2.1cm}
    >{\raggedright\arraybackslash}p{3.0cm}
    >{\raggedright\arraybackslash}p{5.6cm}
    >{\raggedright\arraybackslash}p{4.2cm}
    @{}
}
\toprule
Baseline
&
Tuned hyperparameters
&
Search grid / candidates
&
Best selected setting
\\
\midrule

Multinomial Regression
&
$C$, \texttt{max\_iter}
&
$C \in \{0.01, 0.1, 1.0, 10.0, 100.0\}$; \newline
\texttt{max\_iter} $= 2000$
&
$C = 1.0$; \newline
\texttt{max\_iter} $= 2000$
\\
\addlinespace

Multilayer Perceptron
&
\texttt{hidden\_layer\_sizes}; \newline
\texttt{alpha}; \newline
\texttt{learning\_rate\_init}; \newline
\texttt{early\_stopping}; \newline
\texttt{max\_iter}
&
\texttt{hidden\_layer\_sizes} $\in \{(32,), (64,), (64, 32)\}$; \newline
\texttt{alpha} $\in \{10^{-4}, 10^{-3}, 10^{-2}\}$; \newline
\texttt{learning\_rate\_init} $\in \{10^{-3}, 10^{-2}\}$; \newline
\texttt{early\_stopping} $=$ True; \newline
\texttt{max\_iter} $= 1000$
&
\texttt{hidden\_layer\_sizes} $= (64,)$; \newline
\texttt{alpha} $= 10^{-4}$; \newline
\texttt{learning\_rate\_init} $= 10^{-3}$; \newline
\texttt{early\_stopping} $=$ True; \newline
\texttt{max\_iter} $= 1000$
\\
\addlinespace

XGBoost
&
\texttt{n\_estimators}; \newline
\texttt{max\_depth}; \newline
\texttt{learning\_rate}; \newline
\texttt{subsample}; \newline
\texttt{colsample\_bytree}
&
\texttt{n\_estimators} $\in \{100, 500, 900, 1400\}$; \newline
\texttt{max\_depth} $\in \{2, 4, 6, 8\}$; \newline
\texttt{learning\_rate} $\in \{0.03, 0.05, 0.1\}$; \newline
\texttt{subsample} $\in \{0.8, 1.0\}$; \newline
\texttt{colsample\_bytree} $\in \{0.8, 1.0\}$
&
\texttt{n\_estimators} $= 1400$; \newline
\texttt{max\_depth} $= 8$; \newline
\texttt{learning\_rate} $= 0.03$; \newline
\texttt{subsample} $= 1.0$; \newline
\texttt{colsample\_bytree} $= 0.8$
\\
\addlinespace

Cox PH
&
\texttt{cox\_step\_size}
&
\texttt{cox\_step\_size} $\in \{0.03, 0.1, 0.3\}$
&
\texttt{cox\_step\_size} $= 0.03$
\\
\addlinespace

Royston-Parmar
&
\texttt{rp\_df}; \newline
\texttt{rp\_time\_scale}
&
\texttt{rp\_df} $\in \{2, 3, 4, 5\}$; \newline
\texttt{rp\_time\_scale} $\in \{30.0, 100.0, 365.0\}$
&
\texttt{rp\_df} $= 3$; \newline
\texttt{rp\_time\_scale} $= 30.0$
\\

\bottomrule
\end{tabular}
\end{table}

\subsection{Choosing Latent Dimension Size}
\label{subsec:choosingK}
\begin{figure}[ht]
    \centering
    \includegraphics[width=\linewidth, trim=0 0.425cm 0 0,clip]{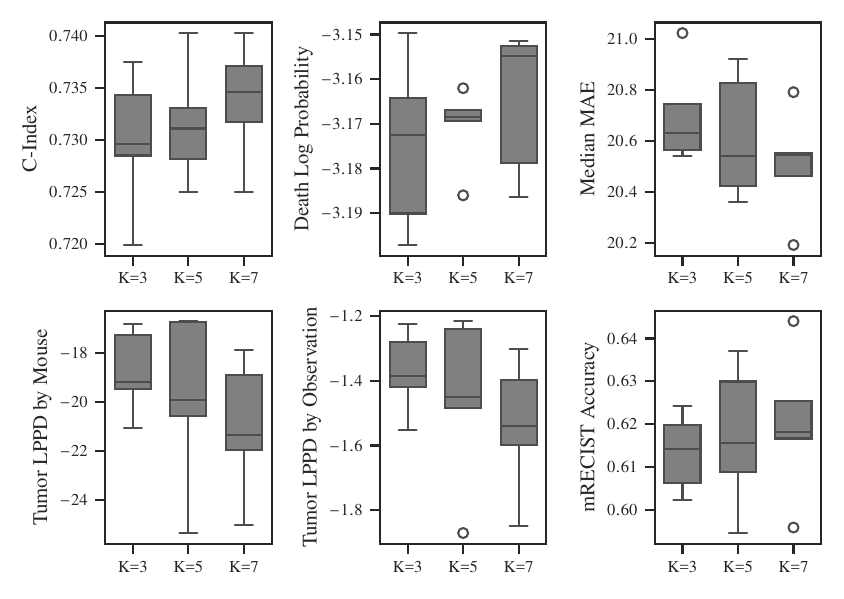}
    \caption{Held-out performance for $K \in \{3,5,7\}$. Survival metrics are similar across the three choices, while held-out tumor LPPD decreases for $K=7$. We therefore use $K=5$ in the main experiments.}
    \label{fig:ChoosingK}
\end{figure}

We quantify performance of \modelname for different values of latent dimension \(K\) on fully held-out data, \(q = 0\). We consider \(K\in\{3,5,7\}\) and hold out 20\% of the drug-PDX combinations as a test set. We repeat this procedure across 10 different train-test splits and calculate the relevant metrics on each split. We report the mean absolute error (MAE) and the average log pointwise predictive density (LPPD)~\citep{Gelman2014}. We complement LPPD with three held-out diagnostics: time-of-death absolute error for the posterior median, concordance index (C-index), and mRECIST accuracy. \Cref{fig:ChoosingK} shows the results of this testing. C-index, LPPD for survival, median MAE, and mRECIST accuracy are all approximately the same across our various values of \(K\). Tumor size LPPD seems to slightly decrease beyond \(K=5\). Ultimately, we choose \(K=5\) because it gives comparable survival performance to other choices while avoiding the apparent decline in tumor predictive density seen at \(K=7\). 

\subsection{Choice of positivity parameterization}
\label{sec:parameter-constraints}

\begin{figure}[ht]
    \centering
    \includegraphics[width=\linewidth, trim=0 1cm 0 0,clip]{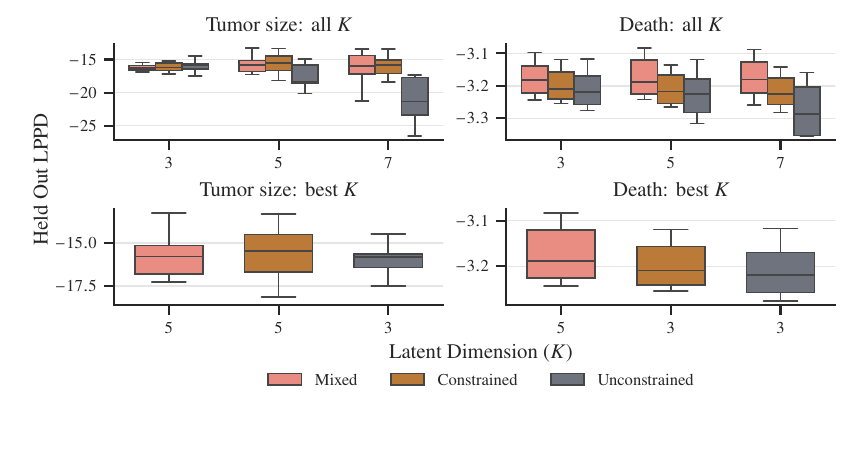}
    \caption{
        Held-out tumor and death LPPD under the mixed, constrained, and log-linked positivity constructions across 10 common train-test splits. The mixed construction yields the highest death LPPD at every fixed \(K\) and after allowing \(K\) to vary, while retaining tumor LPPD comparable to alternative constructions.
    }
    \label{fig:parameter-constraints}
\end{figure}

Both the tumor trajectory and baseline hazard require positive spline coefficients. Because the B-spline basis functions are nonnegative, positivity can be imposed either by constraining the latent factors directly or by applying an exponential link after reconstructing the log coefficient tensor. Let
\[
    \mathcal{T}_{\ell,j,c}(u,v,w)
    \defeq
    \sum_{k=1}^K u_{\ell,k}v_{j,k}w_{c,k}.
\]
We compare three constructions. The \emph{constrained} construction sets both the tumor and baseline hazard coefficients equal to \(\mathcal{T}_{\ell,j,c}\) with nonnegative factors. The \emph{unconstrained} construction exponentiates \(\mathcal{T}_{\ell,j,c}\) in both submodels, allowing the underlying factors to be unconstrained. The \emph{mixed} construction used by \modelname directly constrains the tumor factors but exponentiates an unconstrained factorization for the baseline hazard coefficients. Positive factors retain their \(\mathrm{Exponential}(1)\) priors, unconstrained factors receive \(\mathcal{N}(0,1)\) priors, and all remaining model components, including the nonnegative convolution coefficients \(\beta_c\), are unchanged. We fit each construction with \(K\in\{3,5,7\}\) using the same 10 train-test splits, spline basis, and fitting procedure. \Cref{fig:parameter-constraints} reports held-out tumor and death LPPD. The upper panels compare the constructions at each fixed \(K\), while the lower panels show the value of \(K\) attaining the highest mean LPPD for each construction and outcome.

The mixed construction consistently provides the highest held-out death LPPD. Tumor prediction is broadly comparable between the mixed and constrained constructions. We find that the fully unconstrained construction deteriorates as \(K\) increases. Therefore, we use the mixed construction in the main analysis as it improves survival prediction without any loss in tumor prediction.

\subsection{Full Generative Model}
\label{ref:fullModel}

The following priors are placed on the model parameters:
\begin{align*}
u_{\ell,k}^{\mathsmaller{(d)}}, v_{j,k}^{\mathsmaller{(d)}}, w_{c,k}^{\mathsmaller{(d)}}
&\sim \normal(0,1), \\
u_{\ell,k}, v_{j,k}, w_{c,k},\eta_c
&\sim \mathrm{Exponential}(1), \\
\beta_c
&\sim
\mathrm{Exponential}
\paren{
10^{c-1}
\int_0^T B_c(a)\,da
}.
\end{align*}
for $\ell=1,\ldots,L$, $j=0,\ldots,J$, $k=1,\ldots,K$, and $c=1,\ldots,C$. Here $B_1,\ldots,B_C$ denote the fixed spline basis functions. Together with the tumor and death submodels, these priors define the full generative model. The interval-observed likelihood is used for the PDXe death records, while the right-censored form is included because the held-out prediction task conditions on survival through a censoring time.

\begin{equation*}
\begin{array}{@{}rcl@{\hspace{3em}}rcl@{}}
\multicolumn{3}{c}{\textbf{Death submodel}}
&
\multicolumn{3}{c}{\textbf{Tumor submodel}}
\\[0.75em]

\theta_{\ell,j,c}^{\mathsmaller{(d)}}
&=&
\displaystyle \exp\paren{\sum_{k=1}^K
u_{\ell,k}^{\mathsmaller{(d)}}
v_{j,k}^{\mathsmaller{(d)}}
w_{c,k}^{\mathsmaller{(d)}}}
&
\theta_{\ell,j,c}
&=&
\displaystyle \sum_{k=1}^K u_{\ell,k}v_{j,k}w_{c,k}
\\[1em]

\beta_{\mathrm{conv}}(a)
&=&
\displaystyle \sum_{c=1}^C \beta_c B_c(a)
&
G_{\ell,j}(t)
&=&
\displaystyle \sum_{c=1}^C \theta_{\ell,j,c}B_c(t)
\\[1em]

\lambda_{\ell,j}^{\mathrm{base}}(t)
&=&
\displaystyle \sum_{c=1}^C
\theta_{\ell,j,c}^{\mathsmaller{(d)}}B_c(t)
&
D(t)
&=&
\displaystyle \sum_{c=1}^C \eta_c B_c(t)
\\[1em]

\lambda_{\ell,j}^{\mathrm{tumor}}(t)
&=&
\displaystyle \int_0^t
\beta_{\mathrm{conv}}(t-s)G_{\ell,j}(s)\,ds
&
y_i^{(m)}
&\sim&
\begin{aligned}[t]
\mathrm{Lognormal}\bigl(
&\mathrm{MEDIAN}=G_{\ell_i,j_i}\paren{\tau_i^{(m)}}, \\[-0.4em]
&\mathrm{SCALE}=D\paren{\tau_i^{(m)}}
\bigr)
\end{aligned}
\\[1em]

\lambda_{\ell,j}(t)
&=&
\lambda_{\ell,j}^{\mathrm{base}}(t)
+
\lambda_{\ell,j}^{\mathrm{tumor}}(t)
&
&&
\\[1em]

S_{\ell,j}(t)
&=&
\exp\paren{-\int_0^t \lambda_{\ell,j}(s)\,ds}
&
&&
\\[1em]

\Pr\paren{a_i < t_i^* \le b_i}
&=&
S_{\ell_i,j_i}\paren{a_i}
-
S_{\ell_i,j_i}\paren{b_i}
&
&&
\\[1em]

\Pr\paren{t_i^* > q_i}
&=&
S_{\ell_i,j_i}\paren{q_i}
&
&&
\end{array}
\end{equation*}

\subsection{All Tumor Size Latent Components}
\label{subsec:AllTumorLatents}
\begin{figure}[h]
\centering
\includegraphics[width=\textwidth,height=0.92\textheight,keepaspectratio]{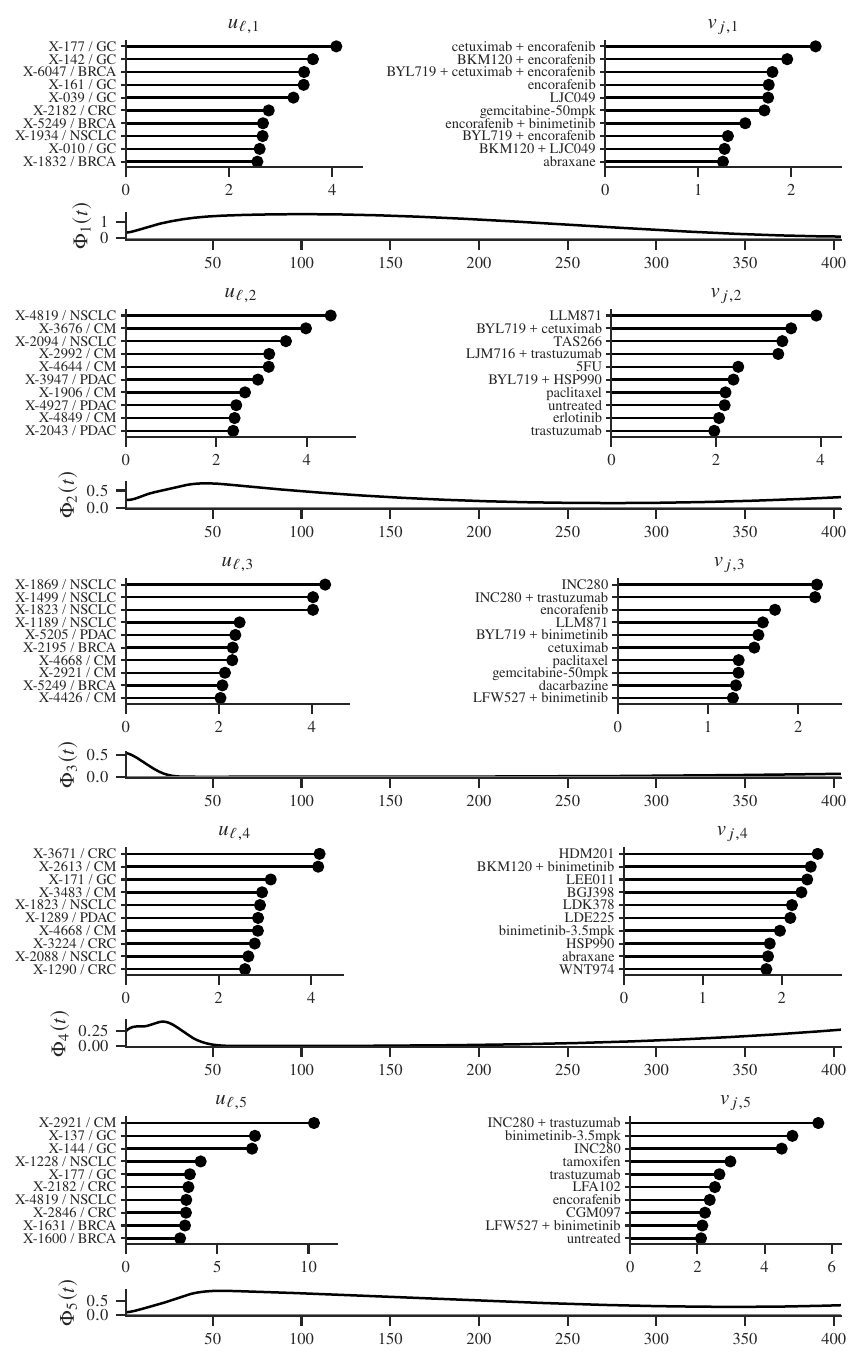}
\caption{Latent tumor components from a single draw of the \(K=5\) model. For each component, the panels display the ten PDX lines with the largest loadings, the ten treatments with the largest loadings, and the corresponding temporal factor.}
\label{fig:tumor-components}
\end{figure}

\end{document}